\documentclass[aps,prb,twocolumn,showpacs,preprintnumbers,amsmath,amssymb,superscriptaddress]{revtex4}
\usepackage{graphicx}
\usepackage{tabularx}
\usepackage{color}
\usepackage{xspace}

\begin{document}
\preprint{PREPRINT (\today)}

\title{ Effect of pressure on the Cu and Pr magnetism in
Nd$_{1-x}$Pr$_x$Ba$_2$Cu$_3$O$_{7-\delta}$ investigated by muon spin rotation}

\author{A.~Maisuradze}\email{alexander.m@physik.uzh.ch}
\affiliation{Physik-Institut der Universit\"{a}t Z\"{u}rich, Winterthurerstrasse 190, CH-8057 Z\"{u}rich, Switzerland}
\affiliation{Laboratory for Muon Spin Spectroscopy, Paul Scherrer Institut, CH-5232 Villigen PSI, Switzerland}
\author{B.~Graneli}
\affiliation{Physik-Institut der Universit\"{a}t Z\"{u}rich, Winterthurerstrasse 190, CH-8057 Z\"{u}rich, Switzerland}
\affiliation{Institute of Theoretical Physics, ETH H\"onggerberg, CH-8093 Z\"urich, Switzerland}
\author{Z.~Guguchia}
\affiliation{Physik-Institut der Universit\"{a}t Z\"{u}rich, Winterthurerstrasse 190, CH-8057 Z\"{u}rich, Switzerland}
\author{A.~Shengelaya}
\affiliation{Department of Physics, Tbilisi State University, Chavchavadze av. 3, GE-0128 Tbilisi, Georgia}
\author{E.~Pomjakushina}
\affiliation{Laboratory for Developments and Methods, Paul Scherrer Institut, CH-5232 Villigen PSI, Switzerland}
\author{K.~Conder}
\affiliation{Laboratory for Developments and Methods, Paul Scherrer Institut, CH-5232 Villigen PSI, Switzerland}
\author{H.~Keller}
\affiliation{Physik-Institut der Universit\"{a}t Z\"{u}rich, Winterthurerstrasse 190, CH-8057 Z\"{u}rich, Switzerland}

\begin{abstract}

The effect of pressure on the copper and praseodymium magnetic order
in the system Nd$_{1-x}$Pr$_x$Ba$_2$Cu$_3$O$_{7-\delta}$ with $x=0.3$, 0.5, 0.7, and 1
was investigated by means of the muon
spin rotation ($\mu$SR) technique. It was found that the effect of pressure
on the N\'eel temperatures of both copper and praseodymium is positive for the whole range of
Pr concentrations ($0.3<x<1$) studied. These findings are in contrast with  a number of previous
reports and clarify some of the puzzles related to the effect of pressure on superconductivity
and magnetism in the praseodymium-substituted
{\it R}$_{1-x}$Pr$_x$Ba$_2$Cu$_3$O$_{7-\delta}$ systems, where {\it R} is a rare earth element.

\end{abstract}
%\pacs{76.75.+i, 74.70.Dd, 74.25.Ha}

\maketitle

\section{Introduction}\label{sec:introduction}

Substitution of rare earth elements ({\it R}) for Y in the cuprate superconductor
YBa$_2$Cu$_3$O$_{7-\delta}$ has been found not to affect the superconducting
properties.\cite{Maple87,Yunhui88,Lin95,Fincher91}
Chemically well prepared samples of {\it R}Ba$_2$Cu$_3$O$_{7-\delta}$ have practically
identical superconducting transition temperatures $T_{\rm c}\simeq 93$~K,\cite{Lin95}
although most of the {\it R}$^{3+}$ elements possess a substantial magnetic moment.\cite{comment1}
Such behavior is explained by a weak hybridization of the {\it R}-layer with the
superconducting CuO$_2$ planes.
However, there are two exceptions: Partially substituting Y with Ce and Pr substantially
reduces $T_{\rm c}$. In {\it R}$_{1-x}$Pr$_x$Ba$_2$Cu$_3$O$_{7-\delta}$
superconductivity is suppressed completely for
$x>x_{\rm cr}$.\cite{Soderholm87Nature,Fincher91,Radousky92rev,Akhavan02rev,Yunhui92,Nieva91,Kebede89}
In the case of Ce the reason for the $T_{\rm c}$ suppression is understood
as a result of its 4+ valence state in contrast to 3+ for
most of the {\it R} elements.\cite{Fincher91}
On the other hand a large number of results obtained with different
techniques point to a 3+ valence state of Pr (see {\it e.g.}
Refs. \onlinecite{Soderholm91,Jostarndt92,Klencsar00}).
Thus, Pr is the exceptional three-valent rare earth element which substantially
influences the superconducting and magnetic properties of
{\it R}$_{1-x}$Pr$_x$Ba$_2$Cu$_3$O$_{7-\delta}$.\cite{Yunhui92, Khasanov08}
Moreover, for a rare earth element {\it R} with a small ionic radius
$R_{\rm i}$ ({\it e.g.} {\it R = }Yb, Lu)
the effect of Pr on $T_{\rm c}$ is relatively small, {\it e.g.} for $x=0.3$ $T_{\rm c}$
is suppressed only by 20\%.\cite{Yunhui92}
On the other hand, for the same Pr concentration $x=0.3$ and for large
 Nd ({\it R = }Nd) superconductivity is suppressed completely,\cite{Yunhui92}
and bulk N\'eel order of the Cu and Pr spins sets in below $T_{\rm N}^{\rm Cu}$ and
$T_{\rm N}^{\rm Pr}$, respectively.
Thus, in contrast to substitution of
Cu by a 3d/4d element which directly and strongly perturbs
the CuO$_2$ planes of the hole-doped cuprates,\cite{Agarval94}
the out of plane perturbation of the superconducting state by Pr is weaker and can be
fine-tuned in a broad range by a hydrostatic or chemical pressure
({\it i.e.} by tuning $R_{\rm i}$).
Understanding of the suppression mechanisms of superconductivity in the
{\it R}$_{1-x}$Pr$_x$Ba$_2$Cu$_3$O$_{7-\delta}$ system, on the other hand,
may help to clarify the pairing mechanisms in the cuprates.\cite{SCinComplSys05, Lee06_RMP}

Soon after the discovery of the unusual behavior of Pr in
the {\it R}$_{1-x}$Pr$_x$Ba$_2$Cu$_3$O$_{7-\delta}$ system first theoretical
explanations of this effect appeared. Fehrenbacher and Rice\cite{Rice93} proposed a
strong hybridization of Pr with the neighboring O ions in the CuO$_2$ plane due to
the large ionic radius of Pr. This idea was further developed by
Liechtenstein and Mazin.\cite{Mazin95,Mazin98} However, there are difficulties
to explain the  chemical and hydrostatic pressure effects in
{\it R}$_{1-x}$Pr$_x$Ba$_2$Cu$_3$O$_{7-\delta}$ by the hybridization model. Namely, a reduction
of $R_{\rm i}$ should increase the hybridization and consequently reduce
$T_{\rm c}$ which is in contrast to experimental observations.\cite{Yunhui92}
Moreover, it was found that PrBa$_2$Cu$_3$O$_{x}$ single crystals prepared by
the traveling-solvent floating-zone technique (TSFZ) or powders prepared by quick quenching
from high temperatures may exhibit a substantial superconducting volume fraction
with $T_{\rm c}$s ranging from zero up to $\sim$85 K.\cite{Zou98,Ye98,Blackstead96}
This finding was later interpreted as a novel realization of superconductivity mediated
by strongly hybridized Pr-O bonds.\cite{Mazin99}

Hydrostatic\cite{Neumeier88,Lin93,Lin96,Lin00} and
chemical\cite{Kebede89,Nieva91,Akhavan02rev,commentPch}
pressure effects (PEs) on $T_{\rm c}$ and the praseodymium  N\'eel
temperature $T_{\rm N}^{\rm Pr}$ in
{\it R}$_{1-x}$Pr$_x$Ba$_2$Cu$_3$O$_{7-\delta}$ are quite controversial.\cite{Radousky92rev,Akhavan02rev}
With increasing ionic radius $R_{\rm i}$ the transition temperature
$T_{\rm c}$ of {\it R}$_{1-x}$Pr$_x$Ba$_2$Cu$_3$O$_{7-\delta}$
gradually reduces and drops to zero (for $x\simeq 0.5$).
Further increase of $R_{\rm i}$ leads
to the onset of N\'eel order of Pr [see Fig. \ref{fig:PhaseDiagr0}(a)].
A similar suppression of $T_{\rm c}$ is observed by the application of hydrostatic
pressure $P$, suggesting an increased localization of carriers with
$P$.\cite{Neumeier88,Lin93,Lin96,Lin00}
This effect  becomes even stronger when
$x$ approaches $x_{\rm cr}$.\cite{comment3} Thus, for $x\simeq x_{\rm cr}$ one
expects an onset of Pr magnetic order and a complete suppression of superconductivity with
increasing $P$  [see Fig. \ref{fig:PhaseDiagr0}(b)].
Surprisingly, though, a rather strong suppression of $T_{\rm N}^{\rm Pr}$ was
reported for PrBa$_2$Cu$_3$O$_{7-\delta}$\cite{Jee88} and PrBa$_2$Cu$_4$O$_{8}$\cite{Weng99} with
$\partial T_{\rm N}^{\rm Pr}$/$\partial P\simeq-10$ K/GPa.
These results indicate a delocalization of the carriers with $P$ and
suggest a complete suppression of $T_{\rm N}^{\rm Pr}$
at $P\simeq 1.7$ GPa ($T_{\rm N}^{\rm Pr}\simeq17$ K for $x=1$).
Furthermore, the superconducting crystals of PrBa$_2$Cu$_3$O$_{x}$
prepared by the TSFZ technique exhibit a positive hydrostatic pressure effect
on $T_{\rm c}$.\cite{Zou98,Ye98}
This observation can be interpreted as an extension of the phase diagram of
$T_{\rm N}^{\rm Pr}$ and T$^{\rm c}$ vs. $P$, provided the samples
prepared by the TSFZ technique experience an extra inhomogeneous chemical pressure caused by
its microstructure which nearly suppresses $T_{\rm N}^{\rm Pr}$ and partly recovers the
superconducting state with a positive pressure effect on $T_{\rm c}$
(see Sec. \ref{sec:discussion} for details).\cite{Akhavan02rev}
 In contrast to these reports, a small and positive PE on $T_{\rm N}^{\rm Pr}$ was detected
by inelastic Neutron scattering (INS) in single crystal of
PrBa$_2$Cu$_3$O$_{x}$.\cite{Lister99}

In order to clarify these controversial experimental results on the pressure effect (PE)
we performed a muon spin rotation ($\mu$SR) study of the PE on
$T_{\rm N}^{\rm Pr}$ and $T_{\rm N}^{\rm Cu}$ in Nd$_{1-x}$Pr$_x$Ba$_2$Cu$_3$O$_{7-\delta}$,
($x=0.3$, 0.5, 0.7, and 1) in the pressure range $ 0<P<2.4$ GPa.
$\mu$SR is a sensitive microscopic probe for studying both magnetic and superconducting
phases of a sample.\cite{musrBookYaouankReotier,Blundell99muSR}
To minimize additional effects of chemical pressure, {\it R} = Nd was chosen,
since Nd and Pr are neighbors in the periodic table of
elements and have nearly identical ionic radii.

We found an increase of $T_{\rm N}^{\rm Pr}$  with increasing
pressure at a rate $\simeq +0.7$ K/GPa for $0.3<x<1$ which is in contrast to
previous reports,\cite{Jee88,Weng99} but consistent with
the negative hydrostatic PE on $T_{\rm c}$  and coincides with the result of PE
on $T_{\rm N}^{\rm Pr}$ determined with INS experiments,\cite{Lister99}
suggesting an increased localization of the carriers with pressure.
We do not observe any superconducting phase for pressures up to 2.3 GPa.
In addition, the copper  N\'eel temperature $T_{\rm N}^{\rm Cu}$ was
found also to increase with  $P$.

The paper is organized as follows: Experimental details and
sample preparation are described in Sec. \ref{sec:ExpDet}.
In Sec. \ref{sec:ResAndAnalysis} we present
results and details of the $\mu$SR data analysis. A discussion
of the effect of hydrostatic pressure on the Cu and Pr  N\'eel ordering in the magnetic part of the
phase diagram is presented in Sec. \ref{sec:discussion}, followed by the conclusions
in Sec. \ref{sec:conclusion}. Some zero- and transverse-field $\mu$SR results for the Co-Ni
alloy (MP35N) used for the nonmagnetic pressure cells are presented in Appendix A.
In Appendix B details of magnetization measurements are given.

\section{Experimental details}\label{sec:ExpDet}

\begin{figure}[!tb]
\includegraphics[width=0.9\linewidth]{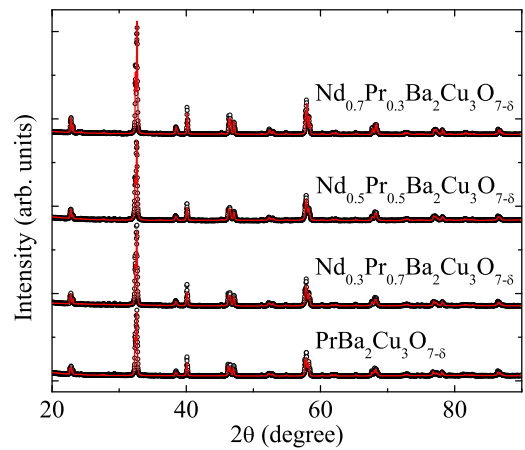}
\caption{(Color online) X-ray powder diffraction patterns for
Nd$_{1-x}$Pr$_x$Ba$_2$Cu$_3$O$_{7-\delta}$  ($x=0.3$, 0.5, 0.7, and 1).
The solid lines are fits to the data. The lattice constants are summarized in
Table \ref{table1}.
\label{fig:FigXray}}
\end{figure}

High quality polycrystalline Nd$_{1-x}$Pr$_x$Ba$_2$Cu$_3$O$_{7-\delta}$ samples with
$x = 0.3$, 0.5, 0.7, and 1
were prepared from Nd$_2$O$_3$, Pr$_6$O$_{11}$, CuO, and BaCO$_3$ of minimum purity of 99.99\%,
using a standard solid state reaction.
Proper amounts of the starting reagents were mixed and calcinated at temperatures
800-920$^\circ$C during at least 150 hours in air with several intermediate
grindings. Finally, the as-prepared samples were oxidized in oxygen atmosphere at a
pressure of 1 bar at 500$^\circ$C. Subsequently, the oxygen contents of the samples
were determined to be close to 7
corresponding to $\delta\simeq 0.02-0.05$.
Powder X-ray diffraction measurements which were carried on a D8 Advance {\it Bruker} AXS
diffractometer using Cu $K_{\alpha}$ radiation indicate that the samples are single phase.
The structural refinements were done using the program {\it FULLPROF}.\cite{FullProf}
The results of these measurements and the refinement profiles are shown in Fig. \ref{fig:FigXray}.
The lattice parameters of the Nd$_{1-x}$Pr$_x$Ba$_2$Cu$_3$O$_{7-\delta}$
samples are in reasonable agreement with previous reports\cite{Kebede89,Ghorbani04}
and are summarized in Table \ref{table1}.

% Table I
\begin{table}[!h]
\caption[~]{ Lattice constants of Nd$_{1-x}$Pr$_x$Ba$_2$Cu$_3$O$_{7-\delta}$ for $x=0.3$, 0.5, 0.7, and 1
obtained from powder X-ray diffraction measurements. }\label{table1}
\begin{center}
\begin{tabularx}{0.95\linewidth}{XXXXXXX }
\hline
\hline
$x$  & $a$ ($\AA$)    &$b$ ($\AA$)   & $c$ ($\AA$)    \\
\hline
0.3  & 3.86078(8)    &3.93271(16)   & 11.76732(42)    \\
0.5  & 3.86191(12)    & 3.91003(14)   & 11.78065(44)    \\
0.7  & 3.86288(10)    &3.90871(11)   & 11.78923(32)    \\
1  & 3.86429(8)    & 3.90624(9)   & 11.79748(23)    \\
\hline
\hline
\end{tabularx}
\end{center}
\end{table}

\begin{figure*}[!tb]
\includegraphics[width=0.99\linewidth,trim=0.4cm 0.4cm 0.4cm 0.4cm, clip=true]{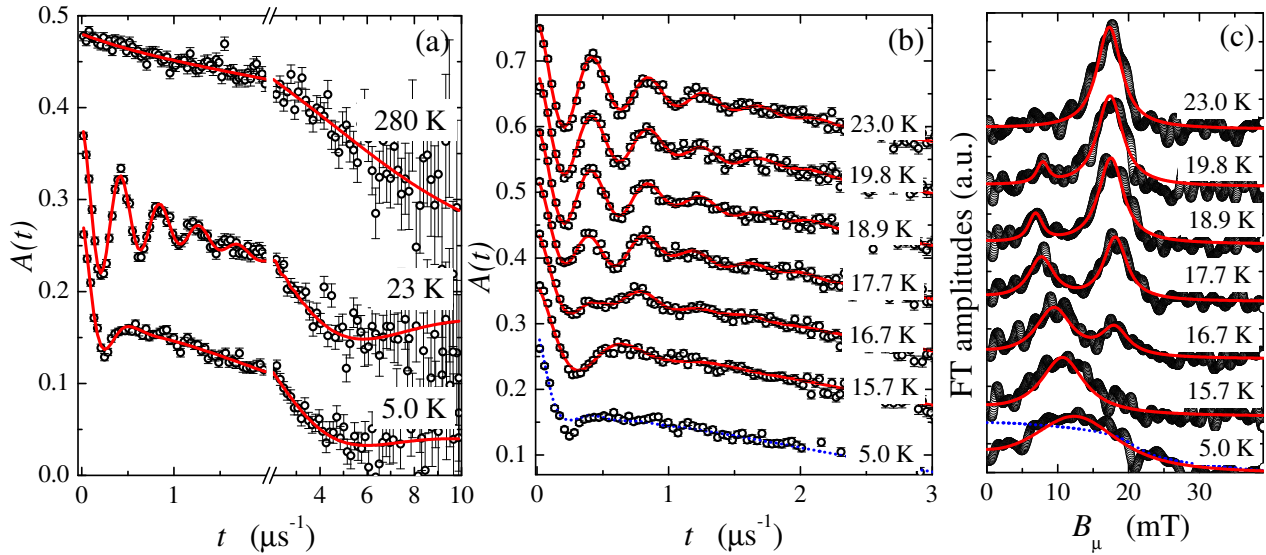}
\caption{(Color online) (a) Zero field $\mu$SR time spectra $A(t)$ of PrBa$_2$Cu$_3$O$_{7-\delta}$
measured at $T = 5$, 23, and 280~K.
The solid lines result from fitting Eqs. (\ref{eq:AsyTot})-(\ref{eq:AsyMagn}) to the data.
For better visualization, the data measured at
23 and 280 K are shifted vertically by 0.1 and 0.2 units, respectively. (b)
The same as in the panel (a) but for $T=5$ K and in the temperature range of 15.7 to 23.0 K.
For better visualization each spectrum is shifted vertically by 0.08 units.
(c) Fourier transform (FT) of the spectra and the corresponding fitted curves shown in the
panels (a) and (b). The dotted lines in panels (b)
and (c) are the best fit to the data and the FT of the Bessel function as described in the text.
\label{fig:Fig1asy}}
\end{figure*}

Zero-field (ZF) $\mu$SR experiments were performed at the $\mu$E1 beam line of
the Paul Scherrer Institute (Villigen, Switzerland) using high energy
muons ($p_\mu \simeq 100$~MeV/c). The samples
were cooled to $\simeq$3\,K, and $\mu$SR spectra were collected in a
sequence with increasing temperature.
Forward and backward positron detectors (relative to the polarization of the incident muon beam)
were employed for the detection of the $\mu$SR asymmetry time spectrum
$A(t)$.\cite{Blundell99muSR,musrBookYaouankReotier}
Typical statistics for a $\mu$SR spectrum were $5\times 10^6$ positron events in the
forward and the backward histograms.
For the $\mu$SR studies pressure was applied with a piston-cylinder type pressure cell ($\mu$SR-PC) of
alloy MP35N using Daphne pressure transmitting oil.\cite{Kamarad04}
The maximum pressure attained was 2.4~GPa at 3~K.
With increasing temperature the pressure increases gradually by $\simeq 0.3$ GPa at $T = 300$ K.\cite{Kamarad04}
Thus, for the applied pressure 2.4~GPa at 3~K we have 2.7 GPa at 300 K.
The actual pressure in the pressure cell was recorded by monitoring the superconducting transition
of a small indium plate mounted in the $\mu$SR-PC next to the sample
(PE on $T_{\rm c}$ of In:
$\partial T_{\rm c}/\partial P = -0.364$~K/GPa).\cite{AndreicaDiss}
In the following we refer to these measured pressures.
Cold pressed samples of cylindrical shape with a size of $5\times 15$ mm (diameter$\times$height)
were used in order to achieve the highest filling factor of the $\mu$SR pressure cell.
The fraction of the muons stopping in the sample was in the range 47-55\%.
The PE studies on the magnetization were performed in a commercial {\it Quantum Design}
(MPMS) XL SQUID magnetometer using a diamond anvil type of pressure cell (DAPC)
with a sample volume of $\simeq 0.1$ mm$^3$ (See Appendix B).

\section{Analysis and results}\label{sec:ResAndAnalysis}

%\begin{widetext}
\begin{figure*}[!htb]
\includegraphics[width=0.32\linewidth]{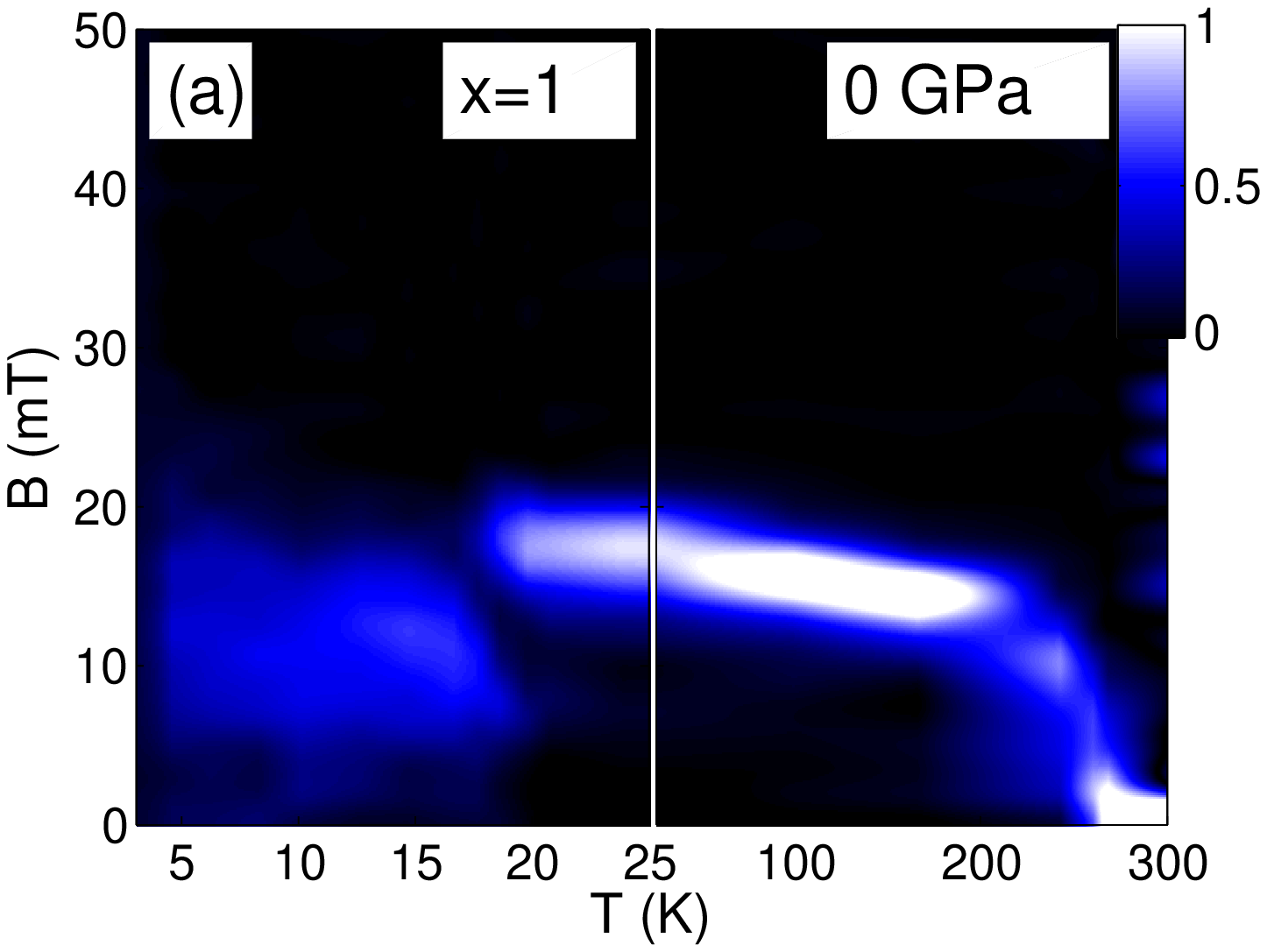}
\includegraphics[width=0.32\linewidth]{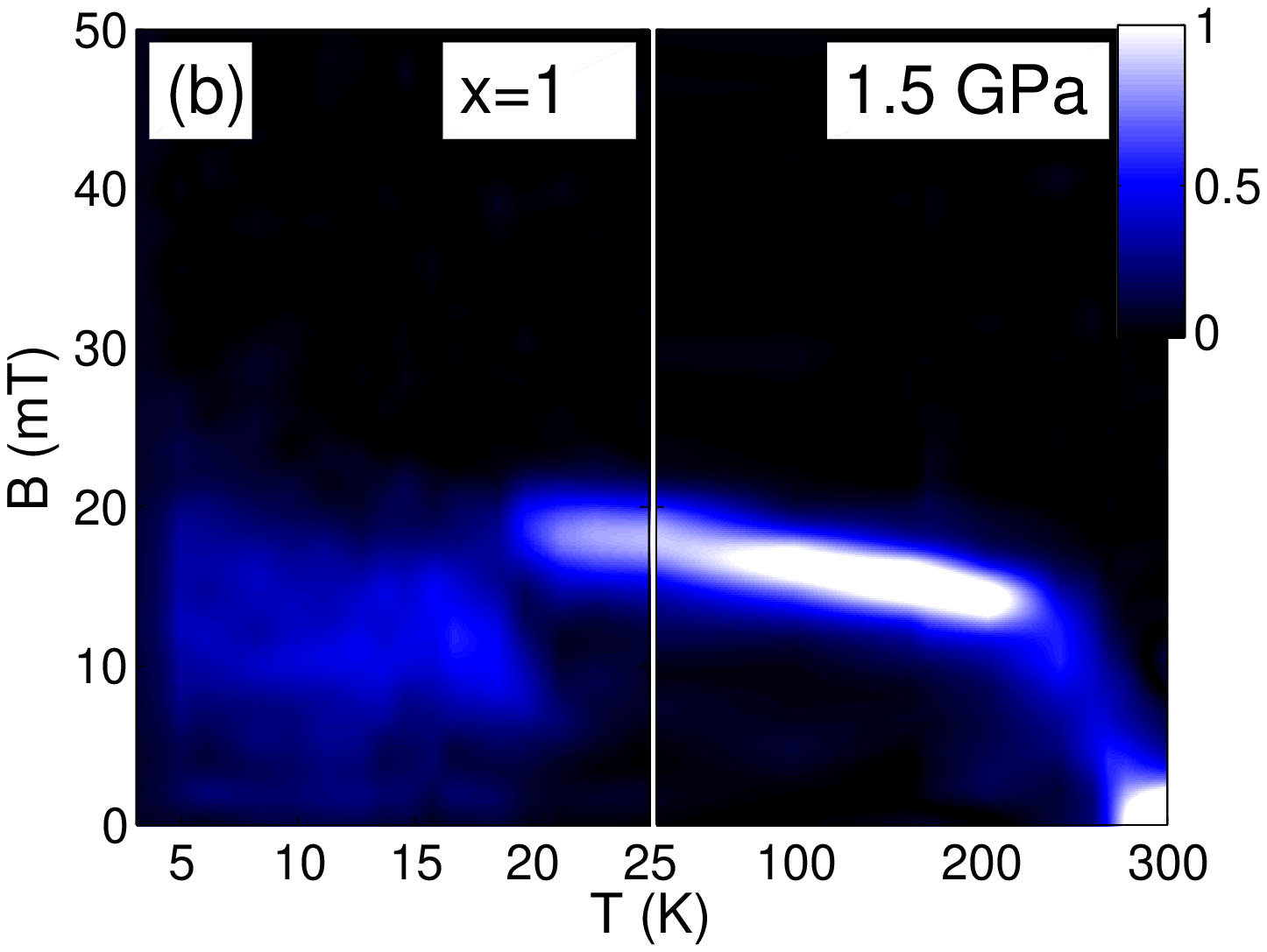}
\includegraphics[width=0.32\linewidth]{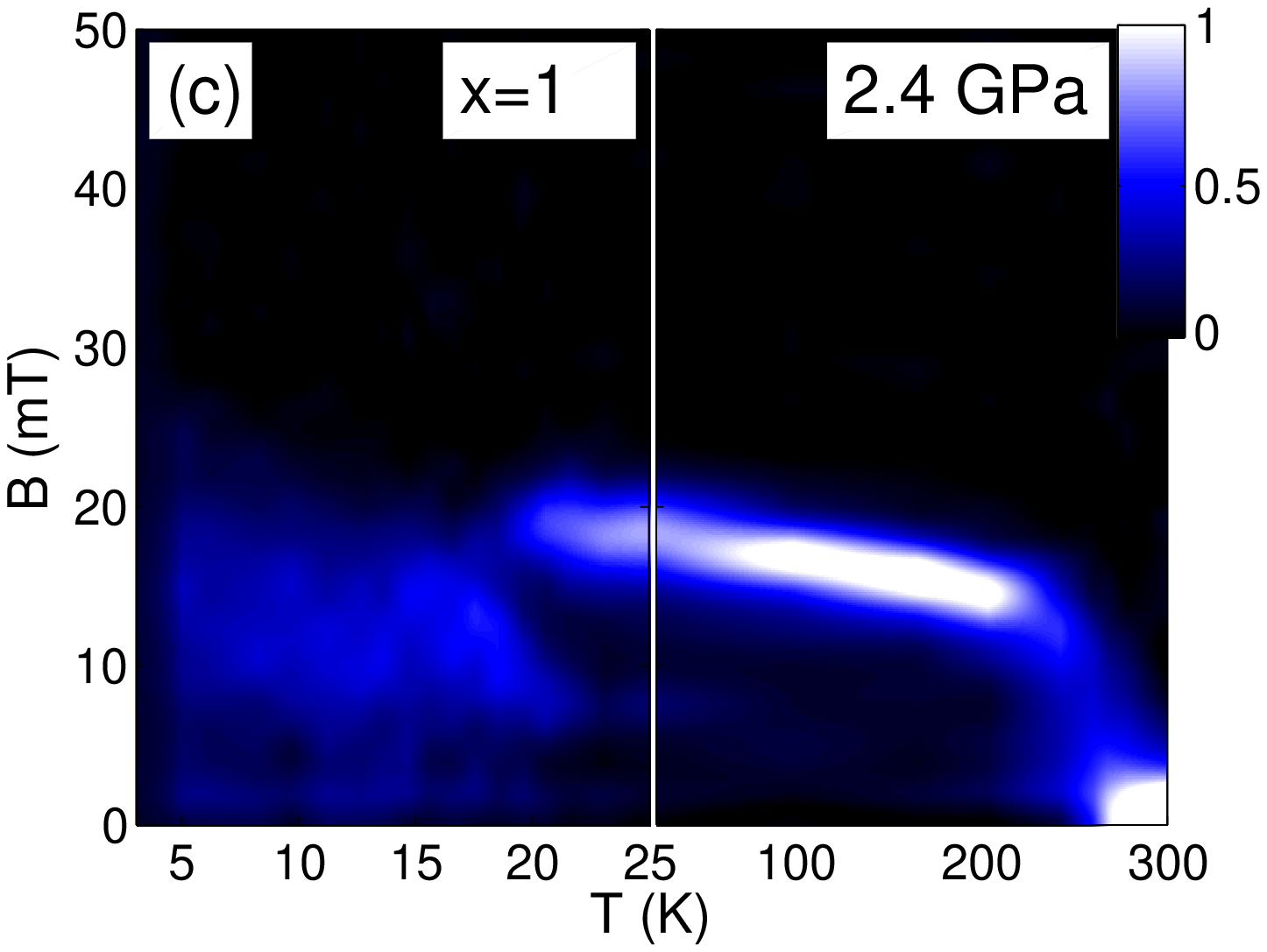}\\
\includegraphics[width=0.32\linewidth]{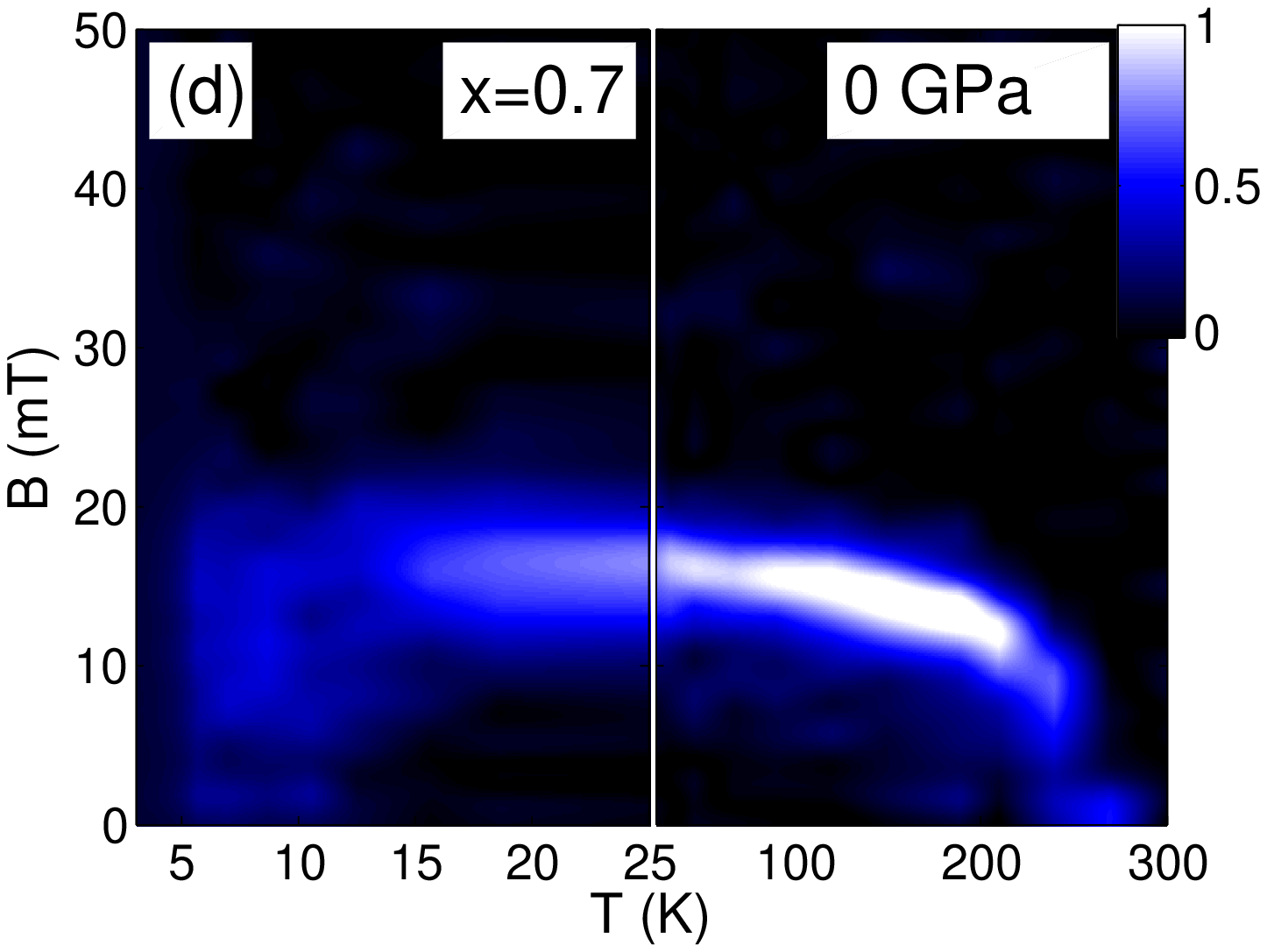}
\includegraphics[width=0.32\linewidth]{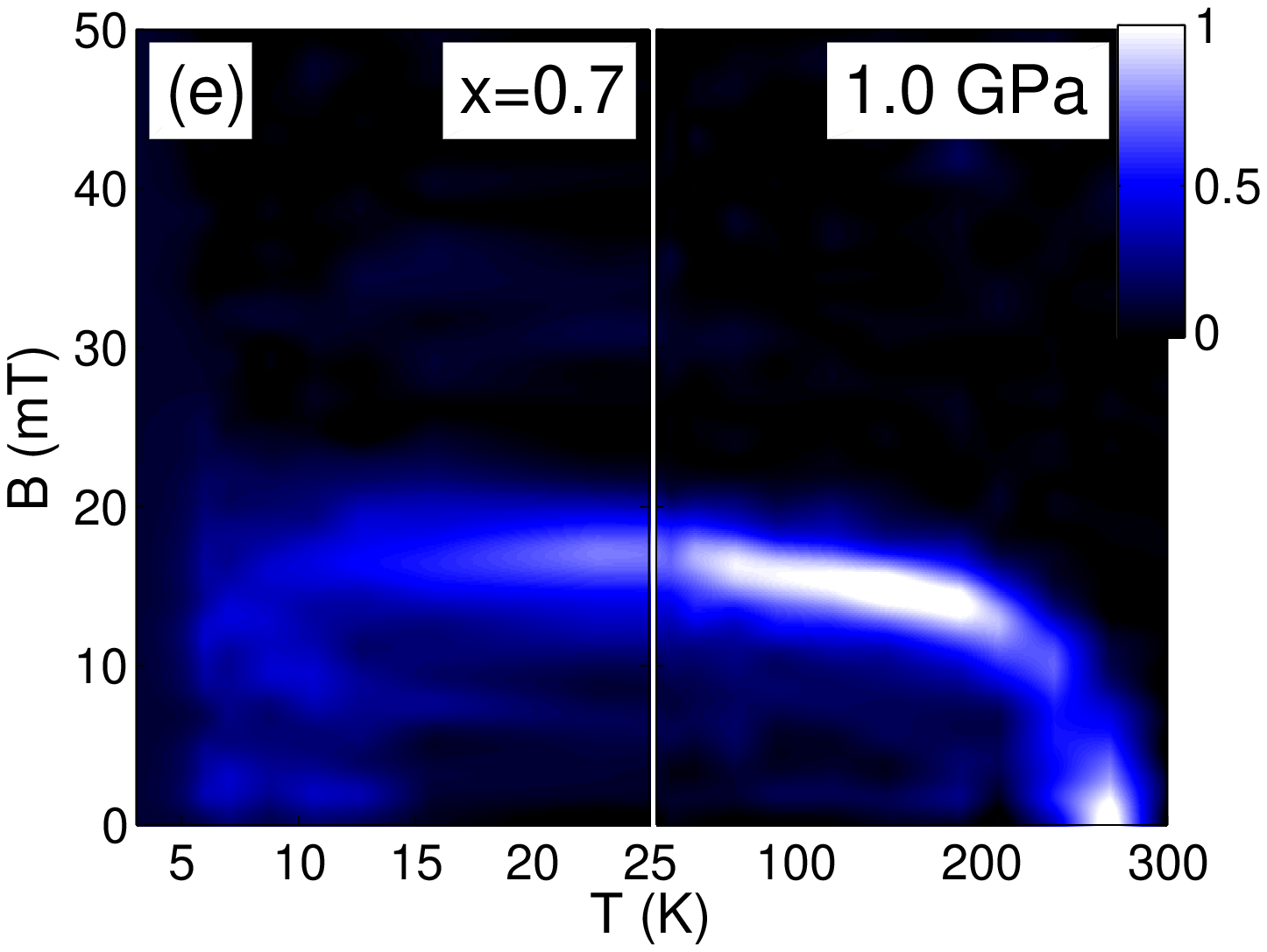}
\includegraphics[width=0.32\linewidth]{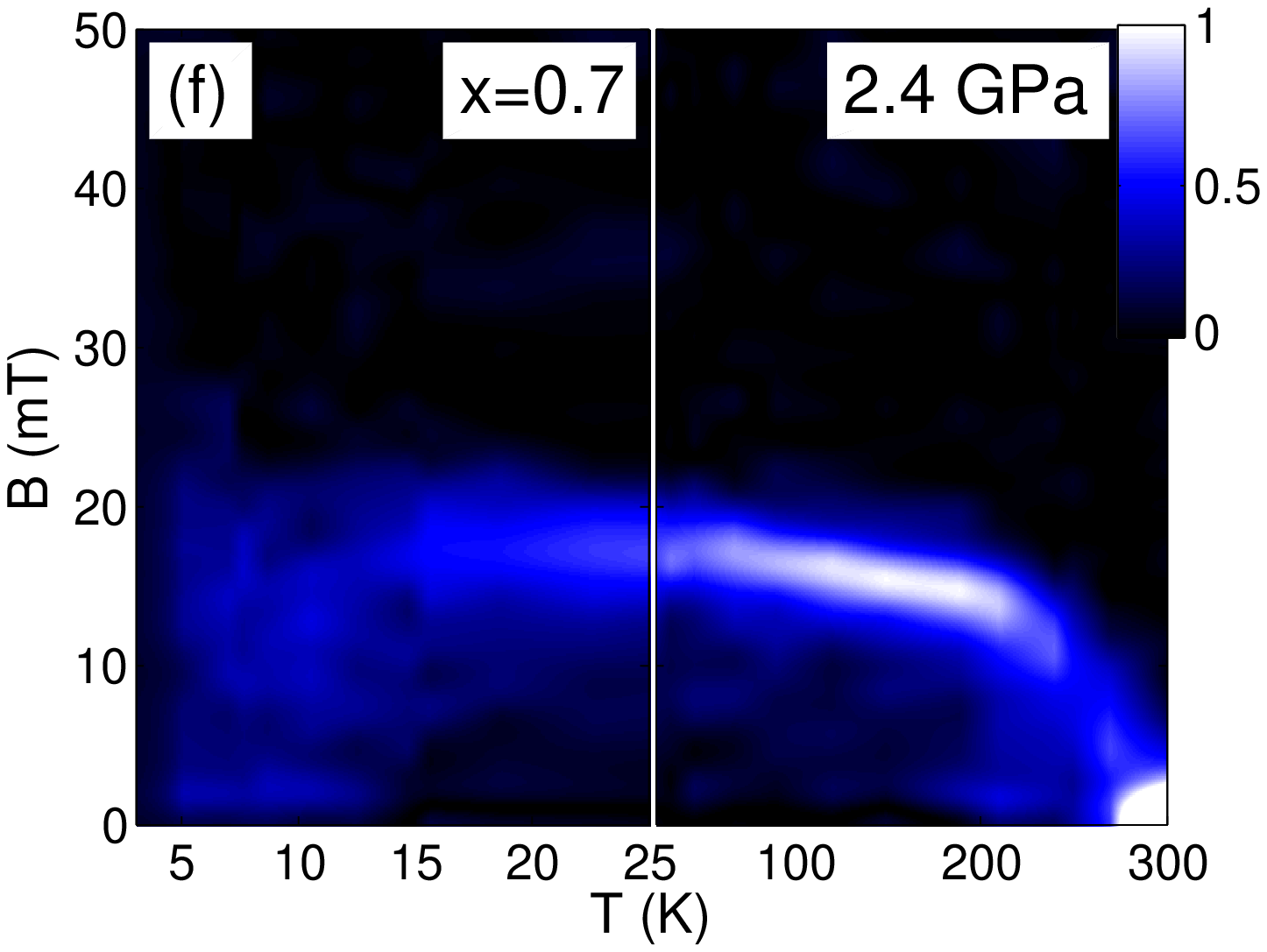}\\
\includegraphics[width=0.32\linewidth]{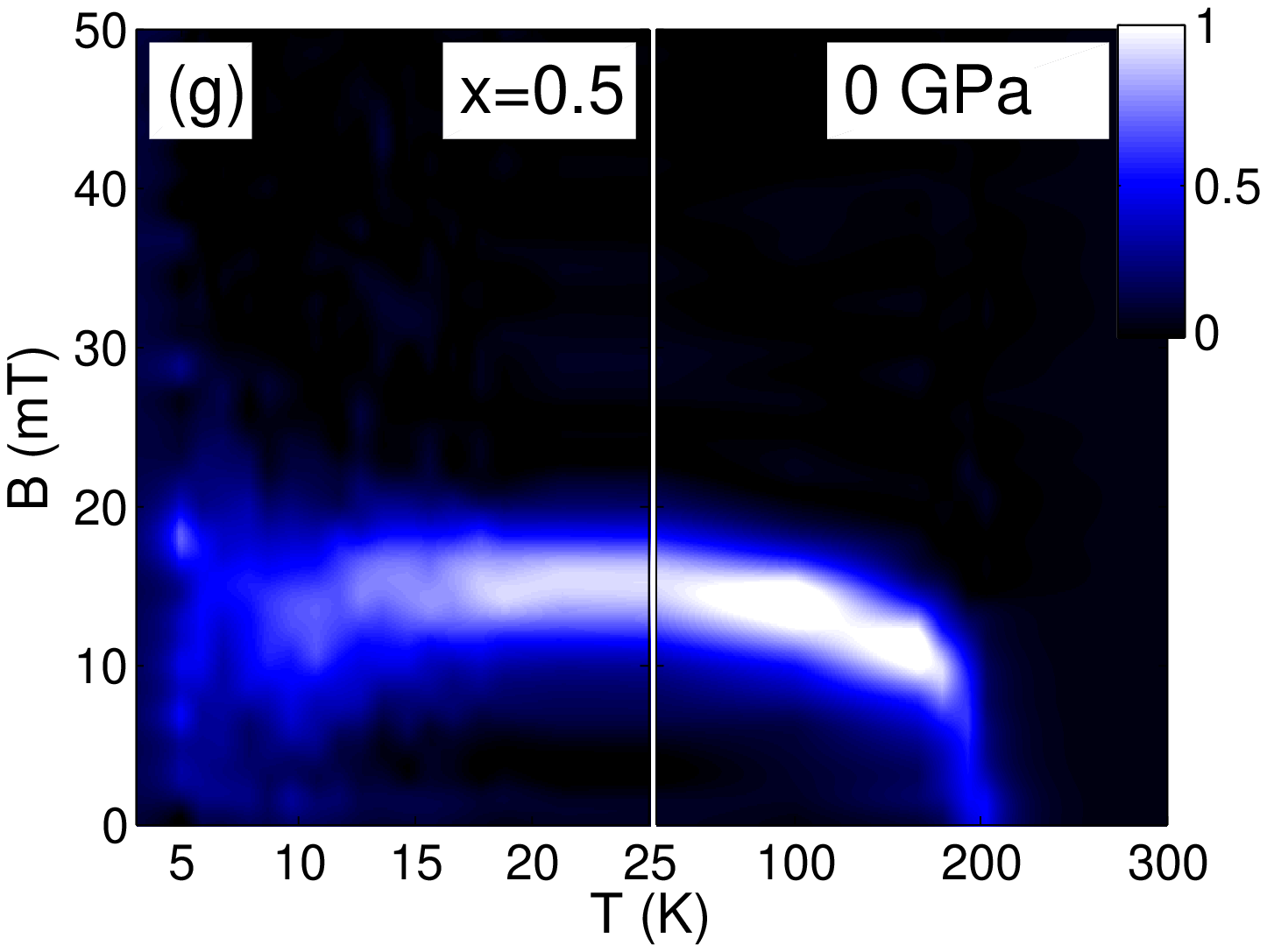}
\includegraphics[width=0.32\linewidth]{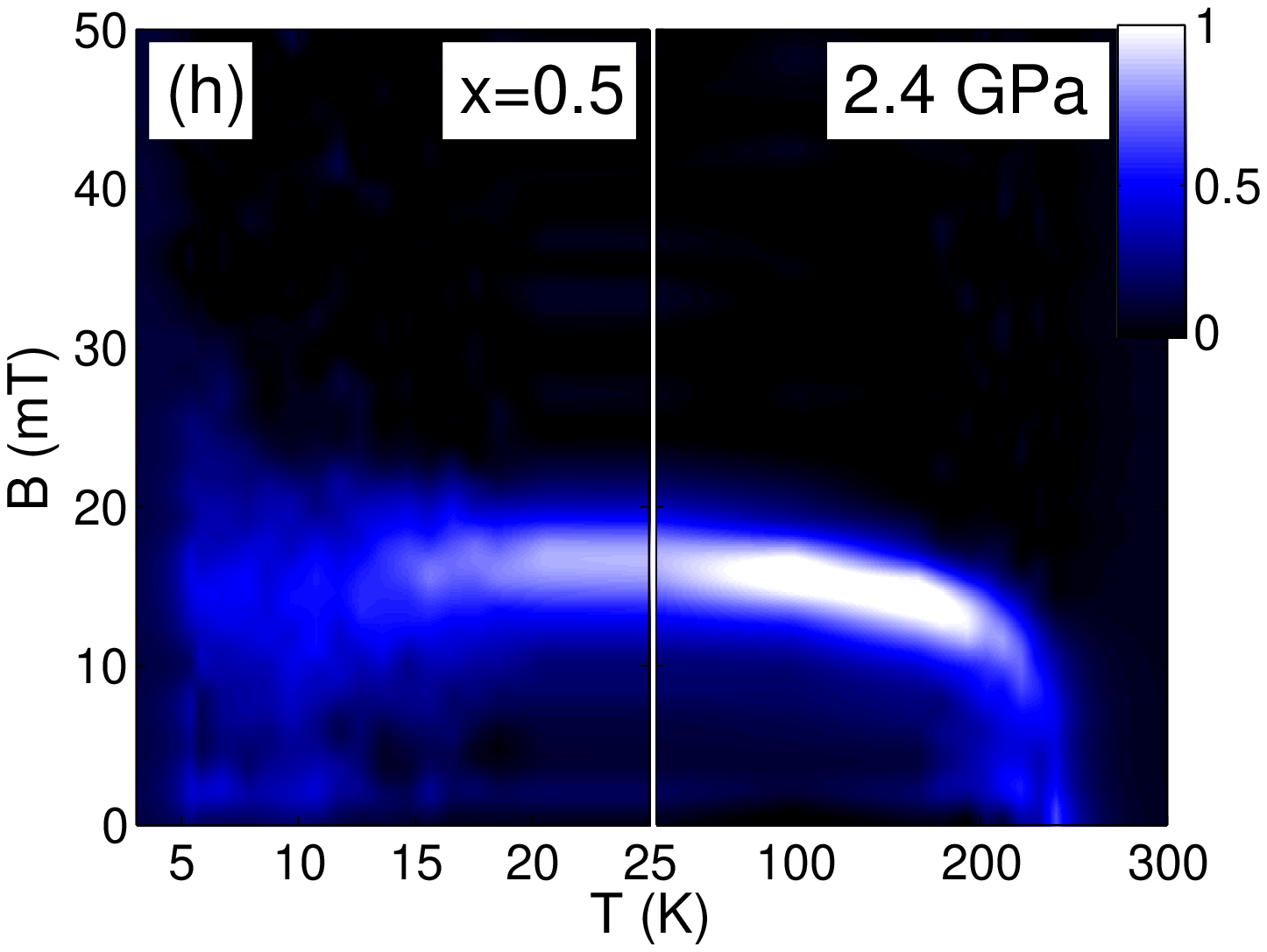}\\
\includegraphics[width=0.32\linewidth]{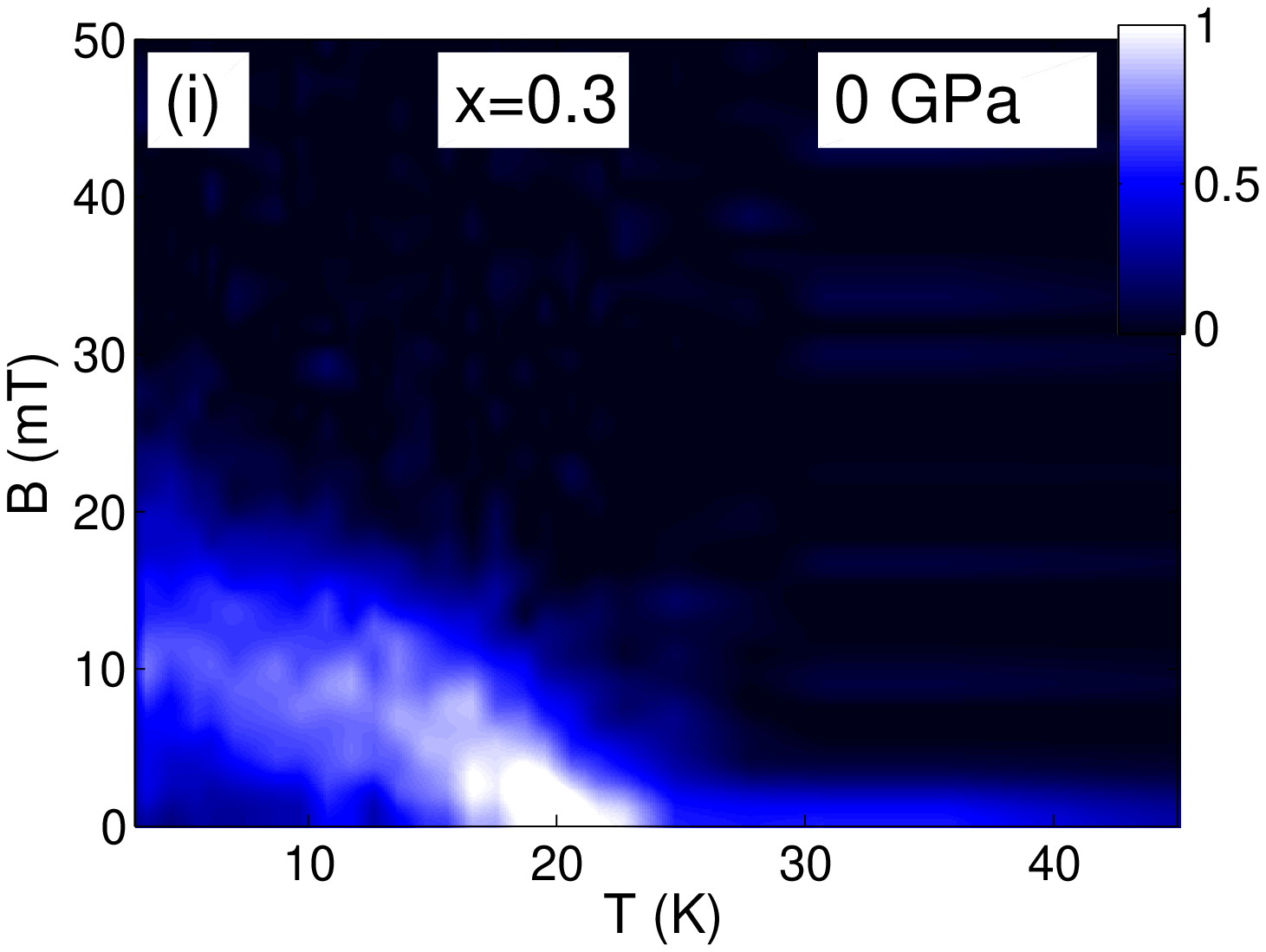}
\includegraphics[width=0.32\linewidth]{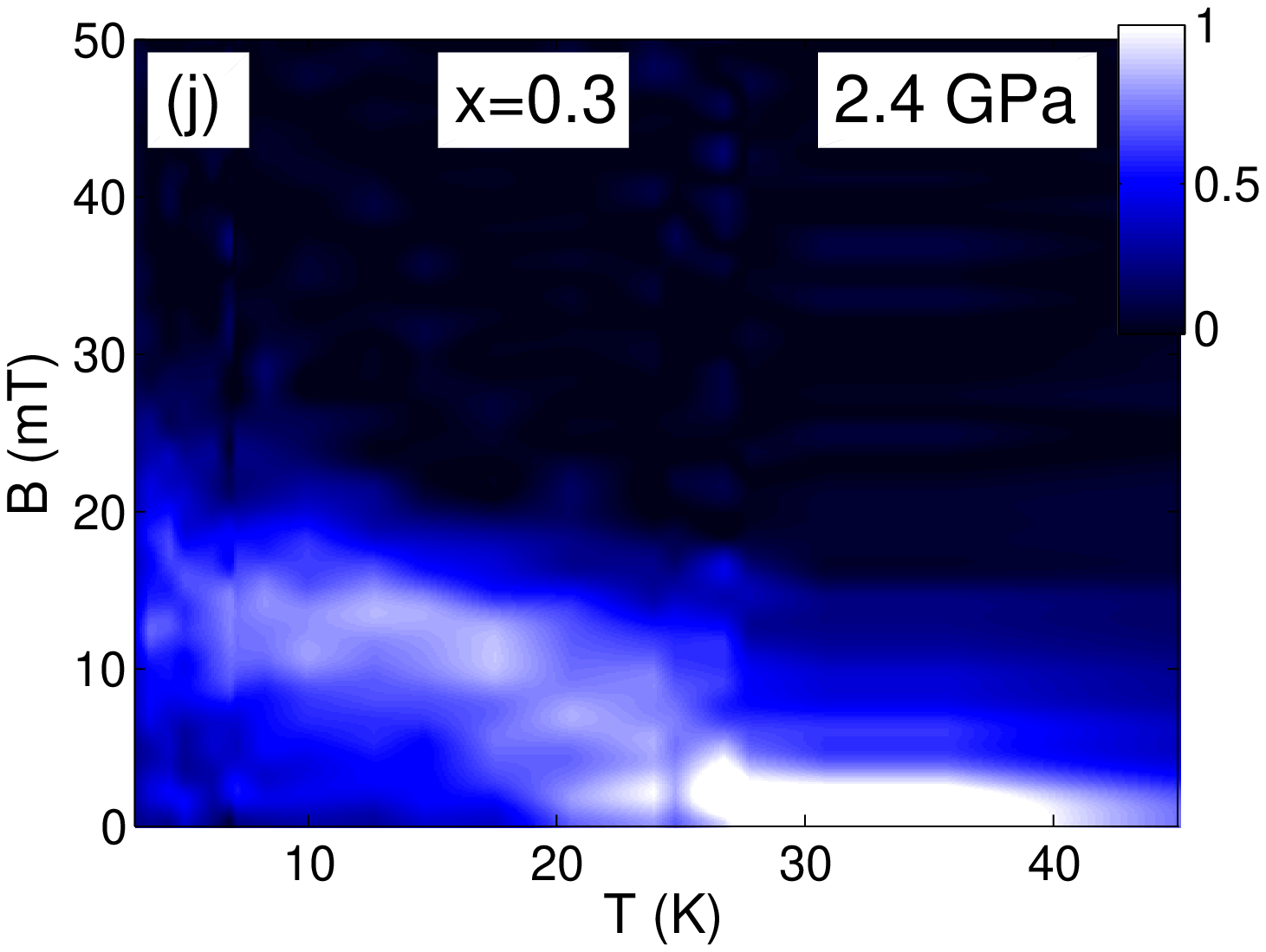}
\caption{(Color online) Fourier transform amplitudes of the oscillating components of the
$\mu$SR time spectra as a function of field and temperature (color maps). Panels (a-c) correspond to
PrBa$_2$Cu$_3$O$_{7-\delta}$ at $P=0$, 1.52, and 2.4 GPa, respectively.
Panels (d-f) correspond to
Pr$_{0.7}$Nd$_{0.3}$Ba$_2$Cu$_3$O$_{7-\delta}$ at $P=0$, 1.0, and 2.4 GPa, respectively;
panels (g) and (h) to
Pr$_{0.5}$Nd$_{0.5}$Ba$_2$Cu$_3$O$_{7-\delta}$ at $P=0$ and 2.4 GPa; panels (i) and (j) to
Pr$_{0.3}$Nd$_{0.7}$Ba$_2$Cu$_3$O$_{7-\delta}$ at $P=0$ and 2.4 GPa.
The amplitudes of the Fourier components were
obtained from the raw data as described in the text.
\label{fig:FigRawFTdata}}
\end{figure*}
%\end{widetext}

The $\mu$SR data are well described in the entire temperature range with the
following equation:\cite{musrBookYaouankReotier}
\begin{equation}\label{eq:AsyTot}
A(t) = A_{\rm pc}P_{\rm pc}(t)+A_{\rm m}P_{\rm m}(t)+A_{\rm pm}\exp(-\lambda_{\rm pm}t)
\end{equation}
The functions $P_{\alpha}(t)$ ($\alpha = $ pc, m) describe the muon polarization functions
of the pressure cell (pc) and the magnetic fraction (m) of the sample, respectively (see below). The last term
corresponds to the asymmetry function of the paramagnetic (pm) fraction of the
sample with a weak muon relaxation rate $\lambda_{\rm pm}\simeq 0.1 - 0.2$ $\mu$s$^{-1}$.
The number of muons stopped in the pressure cell as well as the magnetic and the paramagnetic
fractions of the sample are proportional to the initial asymmetries
$A_{\alpha}$ ($\alpha = $ pc, m, and pm).
The total initial asymmetry $A_{\rm T}=A_{\rm pc}+A_{\rm m}+A_{\rm pm}=0.28$ and
$A_{\rm pc}$ are temperature independent constants.
Typically, the fraction of muons stopped in the sample was $F_{\mu}=(A_{\rm m}+A_{\rm pm})/A_{\rm T}=0.53$.
The value of $F_{\mu}$ depends on the sample position and geometry with respect to
muon beam\cite{AndreicaDiss,Maisuradze11} and is best determined in the temperature
region $25-50$ K, where the samples ($x=1,$ 0.7, and 0.5) exhibit a well developed oscillating
$\mu$SR time spectrum with a slow muon relaxation rate (see Fig. \ref{fig:Fig1asy}).

The muon polarization function of the pressure cell  is described as follows
(see also Appendix A):\cite{AndreicaDiss}
\begin{equation}\label{eq:PcellZF}
P_{\rm pc}(t) = \left[\frac{1}{3}+\frac{2}{3}(1-\sigma_{\rm pc}^2t^2)\exp\left(
{-\frac{1}{2}\sigma_{\rm pc}^2t^2}\right)\right]\exp(-\lambda_{\rm pc}t)
\end{equation}
The relaxation rates of the pressure cell $\sigma_{\rm pc}$
and $\lambda_{\rm pc}$ are given in the Appendix A.

The effect of pressure on the sample is reflected in the second and the third terms of
Eq.~(\ref{eq:AsyTot}). The second term has the form of the standard $\mu$SR depolarization function for a
polycrystalline sample with internal field(s):\cite{Cooke90,Riseman90,musrBookYaouankReotier}
\begin{equation}\label{eq:AsyMagn}
P_{\rm m}(t) = \sum_{i=1}^NF^{\rm m}_i\left(\frac{1}{3}+\frac{2}{3}\cos(\gamma_{\mu}B_it)
e^{-\lambda_{{\rm T},i}t}\right)e^{-\lambda_{\rm L}t}.
\end{equation}
Here, $F^m_i$ represents the fraction of the signal related to the internal field
$B_i$, $N=1$ or 2 (see below), and $\gamma_{\mu}=2\pi\times 135.53$ MHz/T is the
gyromagnetic ratio of the muon. The parameters
$\lambda_{{\rm T},i}$ and $\lambda_{\rm L}$ are the transverse and
longitudinal muon relaxation rates of the $i$-th component.
The longitudinal relaxation rate $\lambda_{{\rm L}}$ is related to  spin dynamics
and is absent in a statically ordered spin system
while $\lambda_{{\rm T},i}$ is proportional to the field inhomogeneity at the muon site.\cite{musrBookYaouankReotier}
The weighting factors 1/3 and 2/3 in Eq.~(\ref{eq:AsyMagn}) originate from
isotropic powder averaging of the $\mu$SR time spectra.
Below we find that $\lambda_{{\rm T},i}\gg \lambda_{\rm L} $.
Results from fitting the $\mu$SR time spectra of  PrBa$_2$Cu$_3$O$_{7-\delta}$
using Eqs. (\ref{eq:AsyTot})-(\ref{eq:AsyMagn}) at $T=5$, 23, and 280 K
are shown in Fig. \ref{fig:Fig1asy}(a). At 5 K a $\mu$SR time spectrum with fast relaxation
was observed, consistent with a magnetic state of the sample.
At 23 K an oscillating signal with slower relaxation is evident, similar
to that observed in previous $\mu$SR studies.\cite{Cooke90,Riseman90}
At 280 K a slowly relaxing non-oscillating signal reflects the paramagnetic state of the sample.
Figure \ref{fig:Fig1asy}(b) shows in more detail the evolution of the $\mu$SR spectra in the
temperature range of 15.7 to 23 K where  Pr-spin ordering occurs,
and Fig. \ref{fig:Fig1asy}(c) presents the corresponding Fourier transform amplitudes
which are described below.

The depolarization function $P_{\rm m}(t)$ given in Eq. (\ref{eq:AsyMagn}) consists of two
separate parts: a fraction $\frac{1}{3}$ that is slowly relaxing and non-oscillating,
and a fraction $\frac{2}{3}$ that is oscillating. Hence, fitting Eqs. (\ref{eq:AsyTot})-(\ref{eq:AsyMagn}) to
the data yields the non-oscillating fraction from the pressure cell, the 1/3
non-oscillating fraction from the sample, and the paramagnetic signal, all of which may be subtracted
from the measured $\mu$SR signal.
The Fourier transform (FT) of the remaining data represents the internal field distribution at
the  muon site(s) of the magnetic part of the sample, provided that the dynamic relaxation
rate is negligible, $\lambda_L\ll\lambda_T$.\cite{musrBookYaouankReotier, Blundell10}
The result of this Fourier transformation for $x=1$ and
$P=0$ in the temperature range 15.7 to 23 K  is shown in Fig. \ref{fig:Fig1asy}(c).
It is noteworthy, that the results are only weakly dependent
on the quality of the fit, since even a FT of
the raw experimental data without subtraction of the non-oscillating components,
yields a very similar result, except for an additional
contribution of the FT amplitudes around zero field.
Figure \ref{fig:FigRawFTdata} shows the FT amplitude color-maps as a function of field and temperature
of the whole experimental data at different $x$ and $P$ providing a quick
and analysis-independent overview of the experimental results for the internal fields probed by $\mu$SR.
For the sample with $x=1$ a sudden decrease
of the internal fields around 16-20 K [see Fig.  \ref{fig:FigRawFTdata}(a-c)]
can clearly be seen, also observed in the previous $\mu$SR studies.\cite{Cooke90,Riseman90}
This drop reflects the  N\'eel transition temperature of the Pr sublattice $T_{\rm N}^{\rm Pr}$.

A more detailed investigation of the data from the sample with $x=1$, however, reveals
two oscillating signals with distinct internal fields in the
temperature range 16-20 K, pointing to a first order magnetic transition [see Fig. \ref{fig:Fig1asy}(c)].
Below $T\simeq 16$ K a single fast relaxation is manifest in the
$\mu$SR signal, which in the analysis emerges as a broad
distribution of internal fields [see Figs. \ref{fig:Fig1asy}(a-c) and \ref{fig:FigRawFTdata}(a-c)].
At higher temperatures ($T>25$ K) again
a single internal field with a narrow distribution is evident, {\it i.e.} a slow transverse
relaxation rate, which gradually drops to zero around
the  N\'eel ordering temperature of the copper sublattice $T_{\rm N}^{\rm Cu}\simeq270$K.
The field at the local muon site  ($B\simeq 18.5$ mT at 23 K) is of the same magnitude as previously
reported\cite{Cooke90,Riseman90} corresponding to a single muon stopping site as
suggested for PrBa$_2$Cu$_3$O$_{7-\delta}$.\cite{Aslanian01}

Early neutron diffraction experiment revealed a simple commensurate antiferromagnetic
order below 17 K in   PrBa$_2$Cu$_3$O$_{7-\delta}$ with
a magnetic moment direction along the c-axis.\cite{Mihalisin89}
Later experiments showed a strong coupling between the Pr- and Cu-spin sublattices with Pr ordering
accompanied by a counterrotation of the Cu antiferromagnetism.\cite{Boothroyd97}
X-ray magnetic scattering point to incommensurate order of the Cu- and the
Pr-spins with a long period of modulation.\cite{Hill98, Hill00}
In Fig. \ref{fig:Fig1asy} we observe a muon depolarization with a small transverse
relaxation rate at $T= 23$ K,
consistent with commensurate antiferromagnetism.\cite{Tranquada88}
However, well below $T_{\rm N}^{\rm Pr}$
the relaxation rate increases strongly, and a broad distribution of
internal fields is evident (see Fig. \ref{fig:Fig1asy}). This broad distribution
indicates a quite disordered magnetic state of the sample.
Strongly coupled Cu- and Pr-spin sublattices
with identical incommensurate modulation periods should give rise to an incommensurate
modulation of the magnetic field at the muon site, leading to a muon depolarization
described by the spherical Bessel function of the first kind $j_0(x)$
[$\cos(\gamma_{\mu}B_it)$ in Eq. (\ref{eq:AsyMagn}) is replaced with $j_0(\gamma_{\mu}B_it)$,
where $j_0(x)=\sin x/x$].\cite{musrBookYaouankReotier}
However,  Eqs.~(\ref{eq:AsyTot})-(\ref{eq:AsyMagn}) describe the data significantly
better than the Bessel function [see Fig. \ref{fig:Fig1asy}(b)].
A Bessel function implies a broad distribution of internal fields with a cut-off\cite{musrBookYaouankReotier}
[see Fig. \ref{fig:Fig1asy}(c)]. But we observe a symmetric Gaussian-like distribution
of fields.

For the samples with $x = 0.3$, 0.5, and 0.7 a
single component of Eq. (\ref{eq:AsyMagn}) suffices
to describe the $\mu$SR time spectra in the whole temperature range
[{\it i.e.} $N=1$ in Eq. (\ref{eq:AsyMagn})].
At the corresponding $T_{\rm N}^{\rm Pr}$ and below, the
mean internal field does not
exhibit a sudden drop [see Figs. \ref{fig:FigRawFTdata}(d)-(j)], but decreases gradually
with a substantial concurrent broadening of the internal field distribution.
At temperatures above $T_{\rm N}^{\rm Pr}$ the mean internal field decreases
with increasing temperature and vanishes above the corresponding $T_{\rm N}^{\rm Cu}$.

In the following sections we describe in more detail the results of the analysis for each of the samples.

\subsection{Results for PrBa$_2$Cu$_3$O$_{7-\delta}$}

\begin{figure}[!ht]
\includegraphics[width=0.9\linewidth,trim=0.4cm 0.4cm 0.4cm 0.4cm, clip=true]{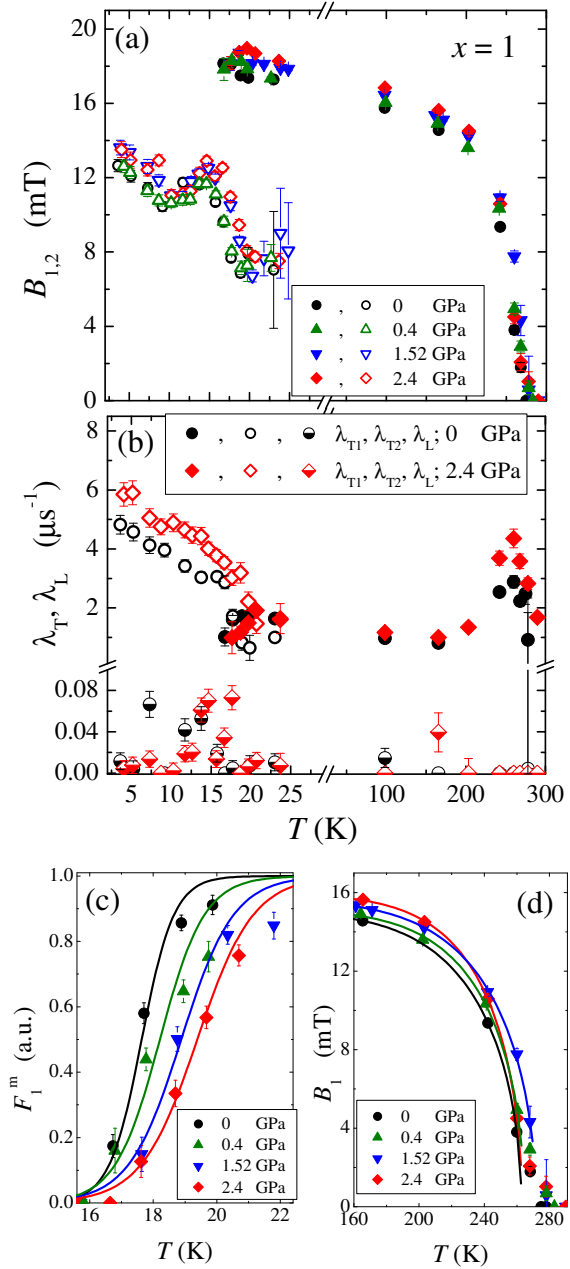} %0.55 0.9
\caption{(Color online) (a) The mean internal fields $B_1$ (full symbols) and $B_2$
(empty symbols) in PrBa$_2$Cu$_3$O$_{7-\delta}$ obtained with
Eqs. (\ref{eq:AsyTot})-(\ref{eq:AsyMagn}) at
$P=0$, 0.4, 1.52, and 2.4 GPa.
(b) Transverse ($\lambda_{T1}$, $\lambda_{T2}$) and longitudinal ($\lambda_L$) relaxation
rates as a function of temperature for $P=0$ and 2.4 GPa (the data for $P=0.4$ and
1.52 GPa are not shown, since they are similar).
(c) Temperature dependence of the magnetic fraction $F^{\rm m}_1$, corresponding to the volume
fraction of the sample with local field $B_1$ (see text).
The solid lines represent best fits to the data using Eq. (\ref{eq:FermiMinus}).
(d) Temperature dependence of $B_1$ in the vicinity of $T_{\rm N}^{\rm Cu}$. The solid
lines are best fits to the data using Eq. (\ref{eq:PowLow}).
\label{fig:Results1p0}}
\end{figure}

The analysis of the $\mu$SR spectra of PrBa$_2$Cu$_3$O$_{7-\delta}$ showed that at zero pressure two
distinct local fields could be identified in the temperature range $16<T<23$ K, where
the onset of magnetic ordering of the Pr sublattice $T_{\rm N}^{\rm Pr}\simeq 17$ K occurs
(see Fig. \ref{fig:FigRawFTdata}). Below 16 K and above 23 K only
a single local field is present.
Indices $i=1, 2$ in Eq. (\ref{eq:AsyMagn}) refer to properties
of the $\mu$SR signals in the high and the
low temperature ranges, respectively. Notations in Eq. (\ref{eq:AsyMagn}) are as follows:
Local fields $B_1$ and $B_2$;
transverse relaxation rates $\lambda_{T1}$ and
$\lambda_{T2}$; weight functions $F_1$ and $F_2$.

The temperature dependence of the mean local field in PrBa$_2$Cu$_3$O$_{7-\delta}$
was obtained by fitting Eqs. (\ref{eq:AsyTot})-(\ref{eq:AsyMagn}) to the data
for pressures $P=0$, 0.4, 1.52, and 2.4 GPa, as shown in
Fig.~\ref{fig:Results1p0}(a) and (d).
The local field $B_2$ has a non-zero value at temperatures above
$T_{\rm N}^{\rm Pr}$ [see Fig. \ref{fig:FigRawFTdata}(a)-(c)] and increases with decreasing temperature.
At temperatures around $T_{\rm N}^{\rm Pr}$
this increase is not monotonic, but exhibits a pronounced peak for all applied pressures
$P=0$, 0.4, 1.52, and 2.4 GPa.

In the high temperature range
field $B_1$ attains its maximum at around 20 K ($B_1\simeq 18$ mT).
With increasing temperature the magnitude of $B_1$ gradually decreases and finally vanishes
at about 270 K, corresponding to the  N\'eel transition temperature of the Cu spin sublattice
$T_{\rm N}^{\rm Cu}$.\cite{Cooke90,Riseman90}
In order to determine $T_{\rm N}^{\rm Cu}$  a power law function was fitted
to the data for $B_1$ above 150 K:
\begin{equation}\label{eq:PowLow}
B(T) = B_0\left( 1 - \left(\frac{T}{T^{\rm Cu}_{\rm N}}\right)^{p}  \right)^{q},
\end{equation}
where the value of $q$ was fixed to 0.5 for all the samples and pressures.
In Table \ref{table2} the values of $T^{\rm Cu}_{\rm N}$ are summarized for all
pressures applied.

The transverse muon relaxation rates $\lambda_{T1}$ and $\lambda_{T2}$,
characterizing the distribution of local fields,\cite{musrBookYaouankReotier}
are shown in Fig. \ref{fig:Results1p0}(b).
For $T<220$ K the relaxation rate $\lambda_{T1}$ is small ($\lambda_{T1}\simeq 1.21$ $\mu$s$^{-1}$)
and at $T\simeq T^{\rm Cu}_{\rm N}$ a peak in $\lambda_{T1}$ is observed.
The relaxation rate $\lambda_{T2}$ increases monotonically
with decreasing temperature below $T^{\rm Pr}_{\rm N}$, which is also evident in the
Fourier amplitudes of the raw experimental data [see Figs. \ref{fig:FigRawFTdata}(a)-(c)].
The amplitudes of the Fourier components gradually decrease below $T^{\rm Pr}_{\rm N}$, whereas the width of
the field distribution increases.
The longitudinal relaxation rate $\lambda_{L}$, characterizing the muon spin
relaxation due to fluctuating magnetic fields is quite small in the whole temperature range
[see Fig. \ref{fig:Results1p0}(b)].

As pointed out above the variation of the local fields $B_1$ and $B_2$ at $T_{\rm N}^{\rm Pr}$ is
discontinuous at all pressures for 16~K~$<T<23$~K, suggesting an inhomogeneous
transition from the Cu to  the Cu+Pr sublattice magnetic order with
decreasing temperature [see Fig. \ref{fig:Fig1asy}(c)].
For a spatially homogeneous transition a gradual broadening of the local field
is expected due to fluctuations of the magnetic order parameter.
The discontinuity is presumably caused either by an electronic inhomogeneity or
by an inhomogeneity of the local chemical pressure.
The corresponding weight functions of the two signals are $F^{\rm m}_i$ ($i=1, 2$)
(with $F^{\rm m}_1+F^{\rm m}_2=1$),
each of which represents a volume fraction of the sample with the local field $B_i$.
Figure~\ref{fig:Results1p0}(c) shows the temperature dependence of $F^{\rm m}_1$ for
$P=0$, 0.4, 1.52, and 2.4 GPa.
Fitting the phenomenological equation:
\begin{equation}\label{eq:FermiMinus}
F(T)=A\left(1-\frac{1}{\exp\left(\frac{T-T_{\rm N}}{\Delta T}\right) + 1}\right) + D
\end{equation}
to the data with $D=0$ and $A=1$, one obtains the solid lines
in Fig.~\ref{fig:Results1p0}(c).
Here, the parameter $A$ describes the maximal value of $F_1^{\rm m}$ while the $\Delta T$
is proportional to width of the magnetic transition.
Note, the second term in the parenthesis resembles the Fermi function.
The N\'eel transition temperature of praseodymium  ($T_{\rm N}^{\rm Pr}=T_{\rm N}$) was
determined from the measured values of $F_1$ by means of Eq.~(\ref{eq:FermiMinus})
in the vicinity of the transition.
Since we have a gradual onset of Pr magnetic order this definition of $T_{\rm N}^{\rm Pr}$
corresponds to the temperature where  half of
the sample volume fraction is in the Pr-ordered state. The width of this
transition is described by the parameter $\Delta T = 0.5$, 0.7, 0.8, and 0.9 K at
$P=0$, 0.4, 1.52, and 2.4 GPa, respectively.
The results for $T_{\rm N}^{\rm Pr}$ are summarized in Table \ref{table2}.

\begin{figure*}[!htb]
\includegraphics[width=0.48\linewidth,trim=0.4cm 0.4cm 0.4cm 0.4cm, clip=true]{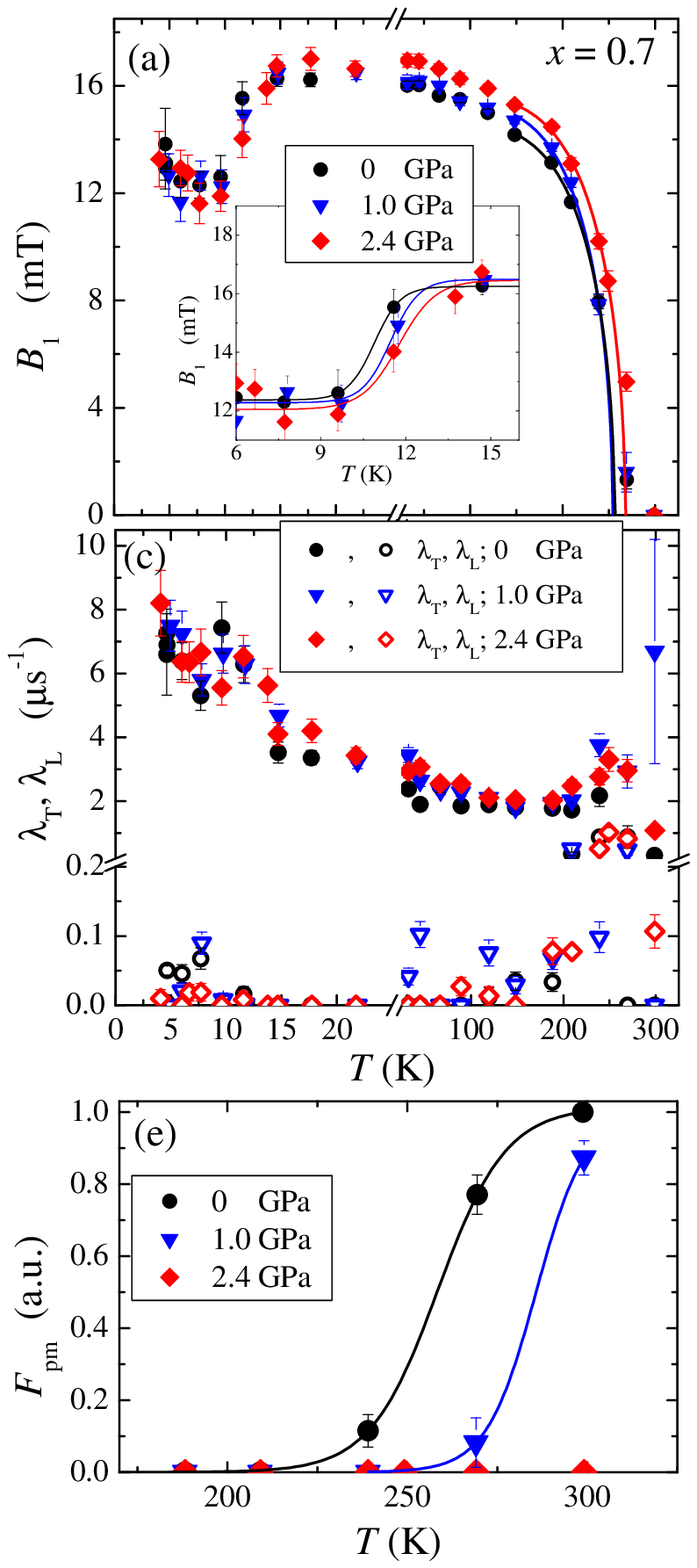}
\includegraphics[width=0.48\linewidth,trim=0.4cm 0.4cm 0.4cm 0.4cm, clip=true]{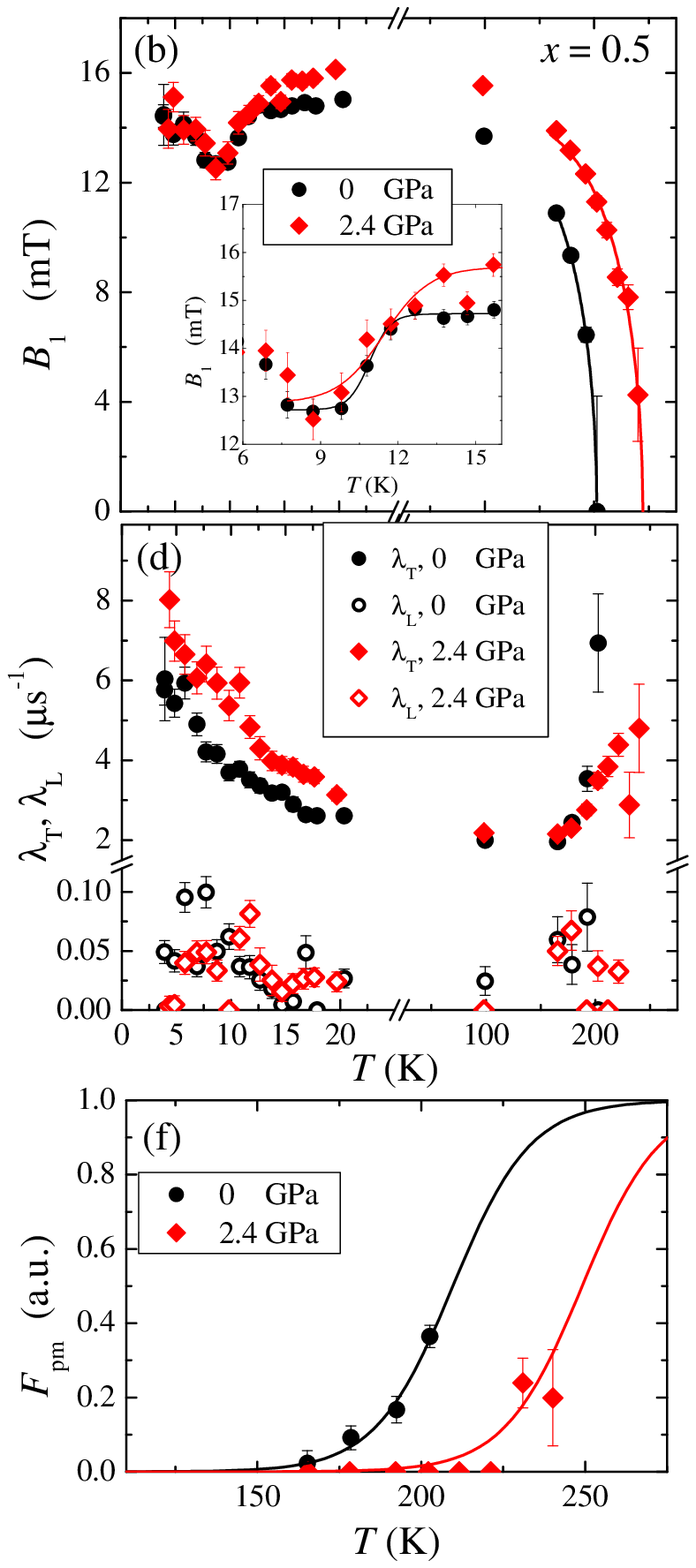}
\caption{(Color online)
(a) The mean internal field $B_1$ in Nd$_{0.3}$Pr$_{0.7}$Ba$_2$Cu$_3$O$_{7-\delta}$
obtained with Eqs. (\ref{eq:AsyTot})-(\ref{eq:AsyMagn}) at applied pressures
$P=0$, 1.0, and 2.4 GPa. (b) The same as panel (a), but for sample
Nd$_{0.5}$Pr$_{0.5}$Ba$_2$Cu$_3$O$_{7-\delta}$ at $P=0$ and 2.4 GPa.
The solid lines in (a) and (b) correspond to the best fits to the data using Eq. (\ref{eq:PowLow}).
The inserts in (a) and (b) show $B_1$ in the vicinity of $T_{\rm N}^{\rm Pr}$
with solid lines representing best fits to the data using Eq. (\ref{eq:FermiMinus}).
(c) and (d) The corresponding transverse ($\lambda_{T}$) and longitudinal ($\lambda_L$) relaxation
rates as a function of temperature for
Nd$_{0.3}$Pr$_{0.7}$Ba$_2$Cu$_3$O$_{7-\delta}$ and
Nd$_{0.5}$Pr$_{0.5}$Ba$_2$Cu$_3$O$_{7-\delta}$, respectively.
(e) and (f) Temperature dependence of the paramagnetic fraction $F_{\rm pm}$
in Nd$_{0.3}$Pr$_{0.7}$Ba$_2$Cu$_3$O$_{7-\delta}$ and
Nd$_{0.5}$Pr$_{0.5}$Ba$_2$Cu$_3$O$_{7-\delta}$, respectively.
The solid lines are results from fitting Eq. (\ref{eq:FermiMinus}) to the data.
\label{fig:Results0p7}}
\end{figure*}

%%%%%%%%%%%%%%%%%%%%%%%%%%%%%%%%%%%%%%%%%%%%%%%%%%%%%%%%%%%%%%%%%%%%%
%%%%%%%%%%%%%%%%%%%%%%%%%%%%%%%%%%%%%%%%%%%%%%%%%%%%%%%%%%%%%%%%%%%%%
\subsection{Results for Nd$_{0.3}$Pr$_{0.7}$Ba$_2$Cu$_3$O$_{7-\delta}$ and Nd$_{0.5}$Pr$_{0.5}$Ba$_2$Cu$_3$O$_{7-\delta}$}

The temperature dependence of the mean local field $B_1$ in
Nd$_{0.3}$Pr$_{0.7}$Ba$_2$Cu$_3$O$_{7-\delta}$
for   $P=0$, 1.0, and 2.4 GPa
and in Nd$_{0.5}$Pr$_{0.5}$Ba$_2$Cu$_3$O$_{7-\delta}$ for  $P=0$ and 2.4 GPa,
obtained by fitting Eqs. (\ref{eq:AsyTot})-(\ref{eq:AsyMagn}) to the data are shown in
Figs.~\ref{fig:Results0p7}(a) and \ref{fig:Results0p7}(b), respectively.

In contrast to  PrBa$_2$Cu$_3$O$_{7-\delta}$,
the $\mu$SR spectra for the two Nd-substituted samples are well described with a
single internal field in the whole temperature range at all pressures applied,
but two distinct temperature regions are evident.
For $T>15$ K the local internal
field gradually decreases with increasing temperature from
$B\simeq16$ mT and vanishes around $T\simeq 260$ K
($T\simeq 200 - 250$ K) in Nd$_{0.3}$Pr$_{0.7}$Ba$_2$Cu$_3$O$_{7-\delta}$
(Nd$_{0.5}$Pr$_{0.5}$Ba$_2$Cu$_3$O$_{7-\delta}$).
This temperature corresponds to $T_{\rm N}^{\rm Cu}$ and was determined by fitting
Eq. (\ref{eq:PowLow}) to the data above 150 K with $q=0.5$.
The corresponding curves are shown as solid lines in Figs. \ref{fig:Results0p7}(a) and (b).
The values of $T_{\rm N}^{\rm Cu}$ for Nd$_{0.3}$Pr$_{0.7}$Ba$_2$Cu$_3$O$_{7-\delta}$
(Nd$_{0.5}$Pr$_{0.5}$Ba$_2$Cu$_3$O$_{7-\delta}$) for
$P=0$, 1.0, and 2.4 GPa ($P=0$ and 2.4 GPa) are given in Table \ref{table2}.

In the low temperature region ($T<15$ K) the mean local field $B_1$
drops for both samples from  $B\simeq16.5$ mT
to $B\simeq12$ mT at $T_{\rm N}^{\rm Pr} \simeq 11$ K.
As is evident from Figs. \ref{fig:FigRawFTdata}(d)-(h)
and \ref{fig:Results0p7}(a) and \ref{fig:Results0p7}(b), the transition is smoother with a
substantial broadening of the field distribution below $T_{\rm N}^{\rm Pr}$.
In order to determine $T_{\rm N}^{\rm Pr}$, the data points in the vicinity of the $T_{\rm N}^{\rm Pr}$
were analyzed with Eq. (\ref{eq:FermiMinus}). The results of this analysis are
represented by the solid lines in
the insert of Fig. \ref{fig:Results0p7}(a) for
Nd$_{0.3}$Pr$_{0.7}$Ba$_2$Cu$_3$O$_{7-\delta}$
and Fig. \ref{fig:Results0p7}(b) for Nd$_{0.5}$Pr$_{0.5}$Ba$_2$Cu$_3$O$_{7-\delta}$.
The corresponding  N\'eel temperatures $T_{\rm N}^{\rm Pr}$ are listed in Table \ref{table2}.

The transverse ($\lambda_{\rm T}$) and longitudinal ($\lambda_{\rm L}$) relaxation
rates for Nd$_{0.3}$Pr$_{0.7}$Ba$_2$Cu$_3$O$_{7-\delta}$
and Nd$_{0.5}$Pr$_{0.5}$Ba$_2$Cu$_3$O$_{7-\delta}$ are shown in Fig. \ref{fig:Results0p7}(c)
and Fig. \ref{fig:Results0p7}(d), respectively.
The random occupation of Nd$^{3+}$ ions at the praseodymium sites gives rise to an
additional disorder of the local fields that is perceived by the muon. Thus, the
transverse relaxation rate $\lambda_{\rm T}$ becomes substantially larger than for
PrBa$_2$Cu$_3$O$_{7-\delta}$ ($x=1$).
The longitudinal relaxation rate $\lambda_{\rm L}$  is small in the whole temperature
range studied, except for  $T\simeq T_{\rm N}^{\rm Cu}$, where $\lambda_{\rm L}$ exhibits a peak.

In contrast to PrBa$_2$Cu$_3$O$_{7-\delta}$, the transition
between the magnetically ordered and the paramagnetic state around
$T_{\rm N}^{\rm Cu}$ is gradual for the samples with $x<1$.
With increasing temperature the mean local field $B_1$ vanishes at $T_{\rm N}^{\rm Cu}$.
Above $T_{\rm N}^{\rm Cu}$, however, a fraction of the $\mu$SR signal still exhibits
a substantial transverse relaxation rate $\lambda_{\rm T}$ with zero mean field, suggesting the presence of
quasi-static and disordered magnetic fields at the muon site.
Below $T_{\rm N}^{\rm Cu}$ a fraction of the sample is in the paramagnetic state
for which the relaxation rate is small: $\lambda_{\rm pm}\simeq 0.1 - 0.2$~$\mu$s$^{-1}$.
Thus, two distinct values of $T_{\rm N}^{\rm Cu}$ may be identified:
(1) the temperature where the mean local field $B_1$ vanishes, and (2) the temperature
where a fraction of one half of the sample volume is in the paramagnetic state.
Close to $T_{\rm N}^{\rm Cu}$ the temperature independent asymmetry of the sample,
$A_{\rm s}=A_{\rm m}+A_{\rm pm}$, may therefore be written as: $A_{\rm m}=A_{\rm s}(1-F_{\rm pm})$
and $A_{\rm pm} = A_{\rm s}F_{\rm pm}$, where the parameter $F_{\rm pm}$ is
the volume fraction of the sample in the paramagnetic state ($0\le F_{\rm pm}\le 1$).
Figures \ref{fig:Results0p7}(e) and \ref{fig:Results0p7}(f) show the temperature dependence
of $F_{\rm pm}$ at different pressures for Nd$_{0.3}$Pr$_{0.7}$Ba$_2$Cu$_3$O$_{7-\delta}$ and
Nd$_{0.5}$Pr$_{0.5}$Ba$_2$Cu$_3$O$_{7-\delta}$, respectively.
The solid lines represent fits of  Eq. (\ref{eq:FermiMinus}) to the data with $A=1$
and $D=0$. The  N\'eel temperature obtained from the temperature dependence
of $F_{\rm pm}$ for various pressures and substitution fractions $x$ is denoted as
$T_{\rm N}^{\rm Cu*}$, and the corresponding values of $T_{\rm N}^{\rm Cu*}$ are
listed in Table \ref{table2}.

\subsection{Results for Nd$_{0.7}$Pr$_{0.3}$Ba$_2$Cu$_3$O$_{7-\delta}$}

\begin{figure}[!htb]
\includegraphics[width=0.9\linewidth,trim=0.4cm 0.4cm 0.4cm 0.4cm, clip=true]{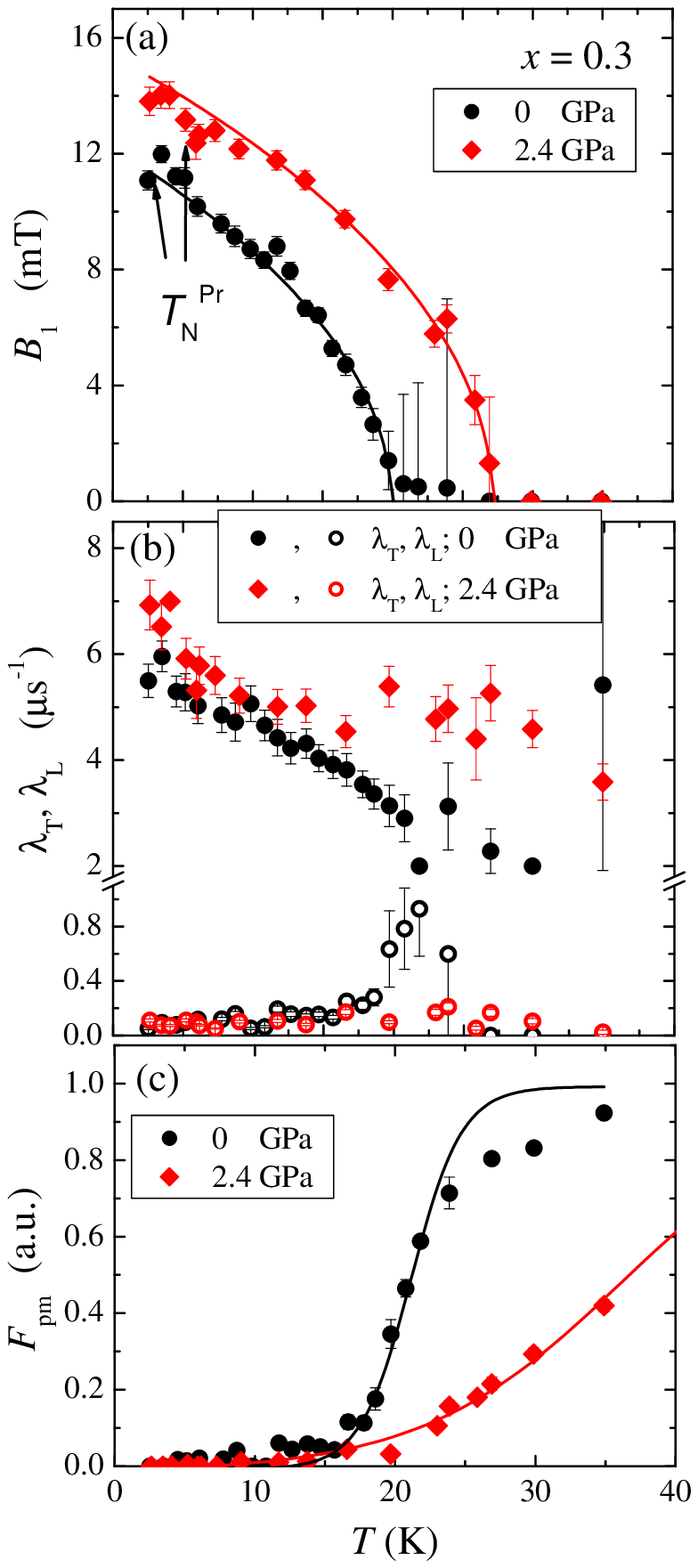} %0.45 0.9
\caption{(Color online)
(a) The mean internal field $B_1$ in
Nd$_{0.7}$Pr$_{0.3}$Ba$_2$Cu$_3$O$_{7-\delta}$
obtained with Eqs. (\ref{eq:AsyTot})-(\ref{eq:AsyMagn}) at
$P=0$ and 2.4 GPa. The solid lines represent best fits to the data using Eq. (\ref{eq:PowLow}).
The arrows indicate possible antiferromagnetic transitions at $T_{\rm N}^{\rm Pr}$.
(b) Corresponding transverse ($\lambda_{\rm T}$) and longitudinal ($\lambda_{\rm L}$) relaxation
rates as a function of temperature for $P=0$ and 2.4 GPa.
(c) Temperature dependence of the paramagnetic fraction $F_{\rm pm}$.
The solid lines are best fits to the data using Eq. (\ref{eq:FermiMinus}).
\label{fig:Results0p3}}
\end{figure}

The temperature dependence of the mean local field $B_1$ in Nd$_{0.7}$Pr$_{0.3}$Ba$_2$Cu$_3$O$_{7-\delta}$
obtained by fitting Eqs. (\ref{eq:AsyTot})-(\ref{eq:AsyMagn}) to the data
for   $P=0$ and 2.4 GPa are shown in Fig.~\ref{fig:Results0p3}(a).
A single component in Eq. (\ref{eq:AsyMagn}) is sufficient to describe the $\mu$SR time
spectra in the whole temperature range.
The local field $B_1$ increases monotonically with decreasing temperature,
but without any pronounced signature at the onset of Pr  N\'eel ordering.
The minor features visible around
3 and 5 K for $P=0$ and 2.4 GPa [shown by the arrows in Fig. \ref{fig:Results0p3}(a)] are possible
signs of Pr ordering, since for the samples with higher
concentration of praseodymium ($x=0.5$ and 0.7), a decrease of $B_1$ at $T_{\rm N}^{\rm Pr}$
is concurrent with an increase of $B_1$ at lower temperatures
[see Figs. \ref{fig:Results0p7}(a) and (b)].
The solid lines in Fig. \ref{fig:Results0p3}(a) represent best fits to the data using
Eq. (\ref{eq:PowLow}) with  $q=0.5$.
The transition temperatures $T_{\rm N}^{\rm Cu}$ obtained from
this analysis for both pressures $P=0$ and 2.4 GPa are summarized
in Table \ref{table2}. The corresponding
transverse ($\lambda_{\rm T}$) and longitudinal ($\lambda_{\rm L}$) relaxation rates are depicted in
Fig. \ref{fig:Results0p3}(b). $\lambda_{\rm T}$ exhibits a monotonic
decrease with increasing temperature, while $\lambda_{\rm L}$ at $P=0$ GPa has a
peak around $T_{\rm N}^{\rm Cu}$. The paramagnetic
volume fraction for both pressures is shown in Fig. \ref{fig:Results0p3}(c).
Similar to the previous case ($x=0.5$ and 0.7) $F_{\rm pm}$
increases gradually with increasing temperature.
The width of this transition, however, gets broad at $P=2.4$ GPa.
The solid lines in Fig. \ref{fig:Results0p3}(c) result from fitting Eq. (\ref{eq:FermiMinus})
to the data with $A=1$ and $D=0$.
The corresponding  N\'eel temperatures
$T_{\rm N}^{\rm Cu*}$ are listed in Table \ref{table2}.

% Table
\begin{table}[!h]
\caption[~]{ N\'eel temperatures for copper ($T_{\rm N}^{\rm Cu}$) and praseodymium
($T_{\rm N}^{\rm Pr}$) in Nd$_{1-x}$Pr$_x$Ba$_2$Cu$_3$O$_{7-\delta}$
($x=0.3$, 0.5, 0.7, and 1.0) at various pressures.
In addition, the values of $T_{\rm N}^{\rm Cu*}$ extracted from the data of the samples
with $x=0.3$, 0.5, and 0.7 are listed (see text for an explanation).}\label{table2}
\begin{center}
\begin{tabularx}{0.95\linewidth}{XXXXXXX }
\hline
\hline
%\vspace{0.1cm}
$x$  &$P$    &$T_{\rm N}^{\rm Pr}$   \hspace{0.3cm}& $T_{\rm N}^{\rm Cu}$ & $T_{\rm N}^{\rm Cu*}$    \\
     &(GPa)     &(K)                   &  (K)    &  (K)   \vspace{0.1cm}  \\
\hline
1&0  &17.6(1)      & 262.8(10)    & --     \\
1&0.4 &18.2(2)      & 264.7(10)   &--      \\
1&1.52&18.9(1)     & 271.6(10)  &--     \\
1&2.4 &19.4(1)     & 263.1(10)     &--  \vspace{0.1cm}    \\
0.7&0  &10.9(2)      & 256(2)     &258(3)    \\
0.7&1.0 &11.5(2)      & 254(2)   &285(3)      \\
0.7&2.4 &11.8(2)    & 267(3)   &$>$300   \vspace{0.1cm}    \\
0.5&0  &10.9(1)      & 202(3)     &209.6(10)    \\
0.5&2.4 &11.4(5)    & 244.4(10)   &249(5)   \vspace{0.1cm}    \\
0.3&0  &3(1)      & 19.9(2)     &21.1(5)    \\
0.3&2.4 &5(1)    & 27.1(2)   &36.2(5)   \vspace{0.1cm}    \\

%\vspace{0.1cm}
\hline
\hline
\end{tabularx}
\end{center}
\end{table}

\section{Discussion}\label{sec:discussion}

\begin{figure*}[!htb]
\includegraphics[width=0.99\linewidth,trim=0.4cm 0.4cm 0.4cm 0.4cm, clip=true]{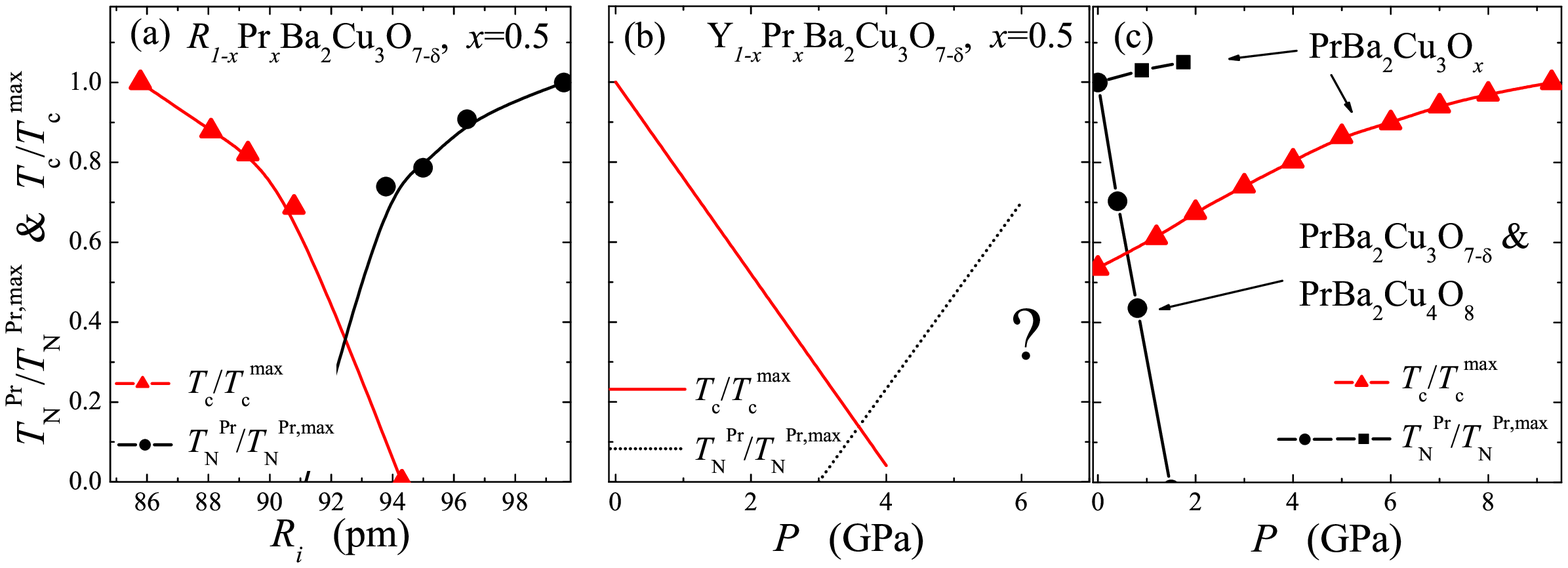}
\caption{(Color online)
(a) $T_{\rm c}/T_{\rm c}^{\rm max}$ and $T_{\rm N}^{\rm Pr}/T_{\rm N}^{\rm Pr,max}$
of {\it R}$_{1-x}$Pr$_{x}$Ba$_2$Cu$_3$O$_{7-\delta}$ as a function of
the rare earth ionic radius $R_{\rm i}$ for  $x=0.5$ with
$T_{\rm c}^{\rm max}=32.6$ K and $T_{\rm N}^{\rm max}=12.1$ K
(the data are from Ref. \onlinecite{Yunhui92}).
(b) $T_{\rm c}/T_{\rm c}^{\rm max}$ for $x=0.5$ and $R=$Y ($T_{\rm c}^{\rm max}=28.9$ K;
the data are from Ref. \onlinecite{Neumeier88}). For all rare earth elements
and $x \lesssim x_{\rm cr}$ the pressure effect on $T_{\rm c}$ is
negative.\cite{Neumeier88,Lin93,Lin96,Lin00} The PE increases in magnitude with
increasing $x$. No studies of the PE on $T_{\rm N}$ for $x<1$ are performed up to now
(indicated by the question mark).
The dotted line shows a possible scenario for the PE on $T_{\rm N}$ suggested by the
results of chemical pressure effect [panel (a)].
(c) $T_{\rm N}^{\rm Pr}/T_{\rm N}^{\rm Pr,max}$ and $T_{\rm c}/T_{\rm c}^{\rm max}$
of PrBa$_2$Cu$_3$O$_{x}$ and PrBa$_2$Cu$_4$O$_{8}$ as
a function of hydrostatic pressure. The results for
$T_{\rm N}$ are from Refs. \onlinecite{Weng99,Jee88}  (circles) with $T_{\rm N}^{\rm max}\simeq 17.5$ K
and Ref. \onlinecite{Lister99} (squares)
with $T_{\rm N}^{\rm max}\simeq 19.6$ K. The pressure effect
data on $T_{\rm c}$ for PrBa$_2$Cu$_3$O$_{x}$ are
from Ref. \onlinecite{Ye98} ($T_{\rm c}^{\rm max}\simeq 105$ K). The solid lines in the panels
(a), (b), and (c) are guides to the eye.
\label{fig:PhaseDiagr0}}
\end{figure*}

Figure \ref{fig:PhaseDiagr0}(a) shows the normalized superconducting
($T_{\rm c}/T_{\rm c}^{\rm max}$) and magnetic
($T_{\rm N}^{\rm Pr}/T_{\rm N}^{\rm Pr,max}$) transition
temperatures in {\it R}$_{1-x}$Pr$_{x}$Ba$_2$Cu$_3$O$_{7-\delta}$ as a
function of ionic radius $R_{\rm i}$ for $x=0.5$ obtained in Ref. \onlinecite{Yunhui92}.
With increasing ionic radius $R_{\rm i}$ of the rare earth element superconductivity
is suppressed. At a sufficiently large $R_{\rm i}$ superconductivity vanishes and Pr-magnetic
order develops, crossing a region of coexistence of both phases.\cite{Yunhui92}
For hydrostatic pressure a negative PE on $T_{\rm c}$ was reported for
{\it R}$_{1-x}$Pr$_{x}$Ba$_2$Cu$_3$O$_{7-\delta}$ at $x \lesssim x_{\rm cr}$
which may be explained by the increased localization of carriers with $P$
[see Fig. \ref{fig:PhaseDiagr0}(b)].\cite{Neumeier88,Lin93,Lin96,Lin00}
The closer $x$ is to the respective $x_{\rm cr}$ the stronger is the PE.
There are no reports on PE on $T_{\rm N}^{\rm Pr}$ for $x<1$ at present.
However, the results of the chemical PE  [see Fig. \ref{fig:PhaseDiagr0}(a)]
and the increased localization of carriers with $P$
suggest a positive PE on $T_{\rm N}^{\rm Pr}$ after complete suppression of superconductivity.
 Zero and a small positive PE on $T_{\rm N}^{\rm Pr}$ was found by transport
and INS experiments, respectively for $x=1$.\cite{Kikuchi99,Lister99}
In contrast, a negative PE on $T_{\rm N}^{\rm Pr}$ was reported
for PrBa$_2$Cu$_3$O$_{7-\delta}$\cite{Jee88} and
PrBa$_2$Cu$_4$O$_{8}$\cite{Weng99} with a large suppression rate:
$\partial T_{\rm N}^{\rm Pr}$/$\partial P\simeq-10$ K/GPa [see Fig. \ref{fig:PhaseDiagr0}(c)].
For $T_{\rm N}^{\rm Pr}=17$ K a linear extrapolation suggests
that for $P\gtrsim 1.7$ GPa the Pr  N\'eel order might be suppressed
completely, and followed by a transition to a superconducting phase.
Indeed,  in PrBa$_2$Cu$_3$O$_{7-\delta}$ thin films,\cite{Blackstead96}
polycrystalline,\cite{Hults98,Cooley98} and single crystal\cite{Zou98,Ye98} samples superconductivity
with an unusually strong and positive effect of hydrostatic pressure\cite{Zou98,Ye98} on
$T_{\rm c}$ has been found.
In Fig. \ref{fig:PhaseDiagr0}(c) we show the results of Ref. \onlinecite{Ye98}.
Figure \ref{fig:PhaseDiagr0}(c) suggests that the observation of superconductivity in
PrBa$_2$Cu$_3$O$_{7-\delta}$ samples might be related to a strong suppression rate
of $T_{\rm N}^{\rm Pr}$ with pressure, provided these superconducting  samples have
an extra initial chemical pressure/stress caused by  a microstructure which partly
suppresses magnetism.

\begin{figure}[!tb]
\includegraphics[width=0.99\linewidth,trim=0.4cm 0.4cm 0.4cm 0.4cm, clip=true]{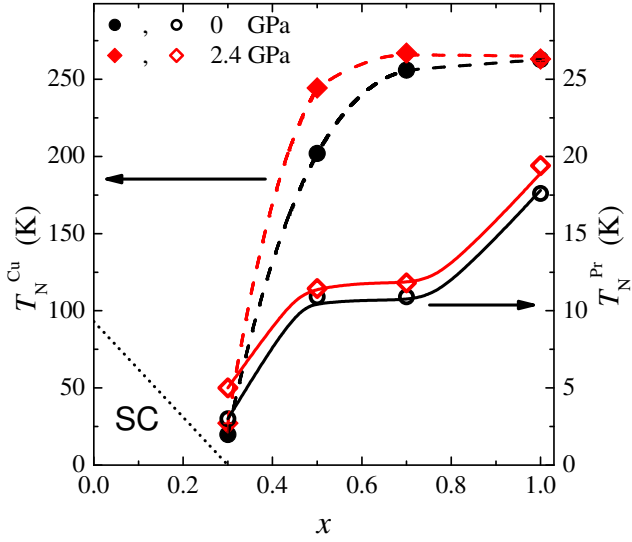}
\caption{(Color online)
Phase diagram of Cu (full symbols) and Pr (empty symbols)
magnetic order in Nd$_{1-x}$Pr$_{x}$Ba$_2$Cu$_3$O$_{7-\delta}$ ($x = 0.3$, 0.5, 0.7, and 1)
at pressures of 0 and 2.4 GPa.
The dotted line indicates the superconducting transition
(see Ref. \onlinecite{Yunhui92}). Solid and dashed lines are guides to the eye.
\label{fig:Summary}}
\end{figure}

\begin{figure}[!tb]
\includegraphics[width=0.99\linewidth,trim=0.4cm 0.4cm 0.4cm 0.4cm, clip=true]{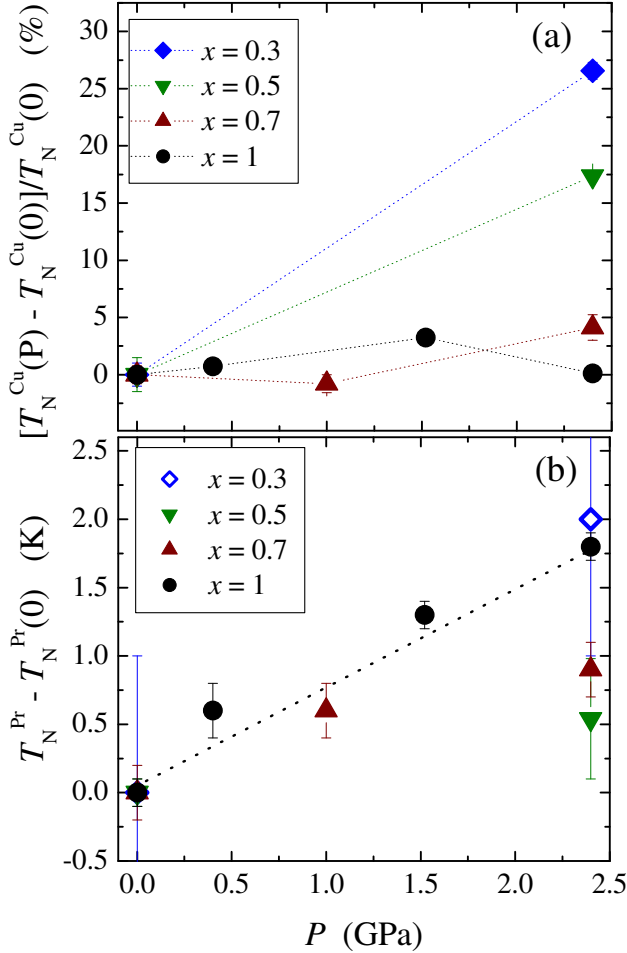}%0.65 0.99
\caption{(Color online) (a) Relative change of the copper N\'eel temperature $T_{\rm N}^{\rm Cu}$
in Nd$_{1-x}$Pr$_{x}$Ba$_2$Cu$_3$O$_{7-\delta}$  as a
function of pressure for $x = 0.3$, 0.5, 0.7, and 1.
(b) Change of the praseodymium N\'eel temperature $T_{\rm N}^{\rm Pr}$ in
Nd$_{1-x}$Pr$_{x}$Ba$_2$Cu$_3$O$_{7-\delta}$ as a function of pressure for
$x=0.3$, 0.5, 0.7, and 1.
The data for $x=0.3$ are shown as empty symbols, since the uncertainty in $T^{\rm Pr}_{\rm N}$
is rather large. The dotted lines are guides to the eye.
\label{fig:Summary2}}
\end{figure}

The magnetic phase diagram for Pr (empty symbols) and Cu (full symbols) in the doping range $0.3<x<1$
is given in Fig.~ \ref{fig:Summary}, showing the  N\'eel transition temperatures $T_{\rm N}^{\rm Pr}$
and $T_{\rm N}^{\rm Cu}$ at pressures 0 and 2.4 GPa.
For $x=1$ the PE on $T^{\rm Cu}_{\rm N}$ is rather small.
The PE increases gradually with  $x$, and attains a maximum for $x\simeq 0.5$.
For $x=0.3$ the effect of hydrostatic pressure
on $T^{\rm Cu}_{\rm N}$ is again small in absolute numbers, but large on a relative scale
[see Fig.~\ref{fig:Summary2}(a)].
Of the several factors determining $T_{\rm N}^{\rm Cu}$ the following deserve attention:
(1) the effective charge carrier concentration $n_h$ in the CuO$_2$ planes, (2) the in-plane and
out-of-plane exchange integrals for the Cu spins.
The former leads to a reduction
of the $T_{\rm N}^{\rm Cu}$ due to disorder introduced by the
localized or hopping carriers in the AF background of the Cu spins.
The latter determines the  N\'eel transition temperature for a system without
charge carriers. Both factors are pressure
dependent. Praseodymium ions localize the carriers in the CuO$_2$ planes, resulting in a reduction of the
effective carrier concentration $n_{h}$ that in turn leads to a suppression of
superconductivity (or increase of $T_{\rm N}^{\rm Cu}$).
This "localization efficiency" strongly depends on chemical pressure,\cite{Yunhui92}
and it is natural to assume that it is also strongly pressure dependent.
On the other hand, with increasing pressure an additional charge transfer to the
CuO$_2$ planes takes place in the YBa$_2$Cu$_3$O$_{x}$ system,\cite{Almasan92,Gupta95,Chen00}
leading to an increase of the maximal superconducting transition temperature
with increasing pressure, or in a different perspective to a decrease of the magnetic transition temperature.
For $x>0.7$ the value of $T_{\rm N}^{\rm Cu}$ only weakly depends on $x$, suggesting
a saturation of the suppression of $n_{h}$ by Pr
(see Fig. \ref{fig:Summary}). Therefore, the effective charge
concentration is too small to influence $T_{\rm N}^{\rm Cu}$ for $x>0.7$ at zero pressure.
$T^{\rm Cu}_{\rm N}$ for $x=1$ increases with increasing pressure up to 1.5 GPa, and then decreases
again for still higher pressures.
Consequently, the weak and nonlinear pressure dependence of $T_{\rm N}^{\rm Cu}$ for $x=1$
is predominantly determined by the PE on the exchange integrals of the Cu spins.
The variation of $T^{\rm Cu}_{\rm N}$
with pressure is substantially stronger for $x=0.5$
than for $x=1$. For $x=0.5$ at zero pressure, $T^{\rm Cu}_{\rm N}$ is reduced
compared to $x=1$ due to the presence of charge carriers.
Thus, the positive PE on $T^{\rm Cu}_{\rm N}$ is substantially influenced by
the negative PE on the effective carrier concentration obviously dominating for $x=0.5$.

The  N\'eel temperature $T_{\rm N}^{\rm Pr}$ at 0 and 2.4 GPa is shown
as a function of Pr concentration $x$ in Fig. \ref{fig:Summary}.
At zero pressure the values of $T^{\rm Pr}_{\rm N}$
are in good agreement with previous reports.\cite{Cooke90,Riseman90,Yunhui92}
In Fig. \ref{fig:Summary2}(b) the variations of
$T^{\rm Pr}_{\rm N}$ for all pressures investigated and all samples are shown.
The PE on $T^{\rm Pr}_{\rm N}$ is positive  ($\simeq +0.7$ K/GPa) for
all the praseodymium concentrations  ($0.3<x<1$).

The present results for the PE on both $T_{\rm N}^{\rm Cu}$ and $T_{\rm N}^{\rm Pr}$ suggest an
increased localization of carriers with $P$ and drastically differ from those
reported previously for PrBa$_2$Cu$_3$O$_{7-\delta}$\cite{Jee88} and
PrBa$_2$Cu$_4$O$_{8}$,\cite{Weng99}
as obtained by magnetization measurements under pressure.
It contrasts also with the positive hydrostatic PE on $T_{\rm c}$
found in superconducting PrBa$_2$Cu$_3$O$_{7-\delta}$
(see Fig. \ref{fig:PhaseDiagr0}).\cite{Zou98,Ye98}
Our magnetization measurements of PrBa$_2$Cu$_3$O$_{7-\delta}$ under hydrostatic
pressures up to 2.3 GPa could not reveal a Meissner fraction (with a precision
better than 1 \%),
confirming absence of superconductivity in the present samples (see Appendix B).
The positive PE on $T^{\rm Pr}_{\rm N}$ and $T^{\rm Cu}_{\rm N}$ found
for $x_{\rm cr} \lesssim x <1$
is, however, in good agreement with the results of INS experiments\cite{Lister99} and is
consistent with the negative  PE on $T_{\rm c}$ reported for
{\it R}$_{1-x}$Pr$_{x}$Ba$_2$Cu$_4$O$_{8}$ at
$x \lesssim x_{\rm cr}$.\cite{Neumeier88,Lin93,Lin96,Lin00}
The suppression of superconductivity and the onset of magnetism with $P$
are in accord with the similar behavior of
$T_{\rm c}$ and T$^{\rm Pr}_{\rm N}$  in
{\it R}$_{1-x}$Pr$_{x}$Ba$_2$Cu$_4$O$_{8}$ as a function of
chemical pressure [see Fig. \ref{fig:PhaseDiagr0}(a)].\cite{Yunhui92}
Our findings are also in agreement with Ref. \onlinecite{Kikuchi99},
where $T^{\rm Pr}_{\rm N}$ was identified from transport measurements
for Y$_{1-x}$Pr$_{x}$Ba$_2$Cu$_4$O$_{8}$, at pressures up to 1 GPa,
and found to be pressure independent.
Therefore, our results strongly suggest an increased localization of the carriers
in  Nd$_{1-x}$Pr$_{x}$Ba$_2$Cu$_3$O$_{7-\delta}$
by Pr$^{3+}$  with the application of  hydrostatic pressure
up to 2.4 GPa, even for $x=1$. Further reliable measurements of the PE on $T^{\rm Pr}_{\rm N}$
are needed to verify the results of Refs.~\onlinecite{Jee88,Weng99}.
We think that a positive PE on $T_{\rm c}$, suggesting delocalization of carriers with
pressure, found in single crystal samples of PrBa$_2$Cu$_3$O$_{7-\delta}$ is
exceptional and is related to details of their microstructure.

\begin{figure*}[!htb]
\includegraphics[width=0.45\linewidth,trim=0.4cm 0.4cm 0.4cm 0.4cm, clip=true]{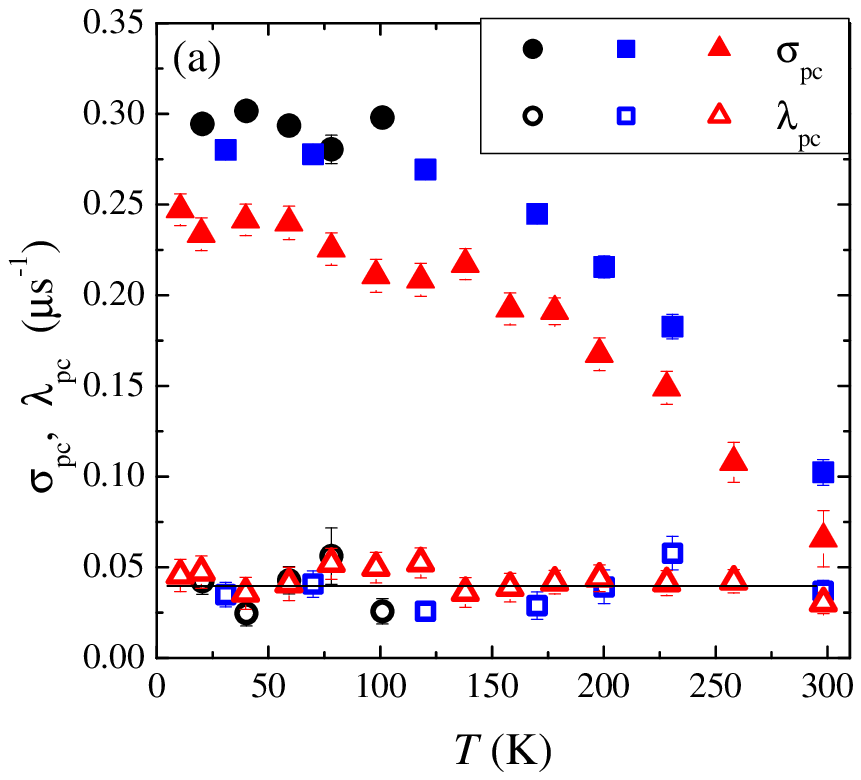}
\includegraphics[width=0.45\linewidth,trim=0.4cm 0.4cm 0.4cm 0.4cm, clip=true]{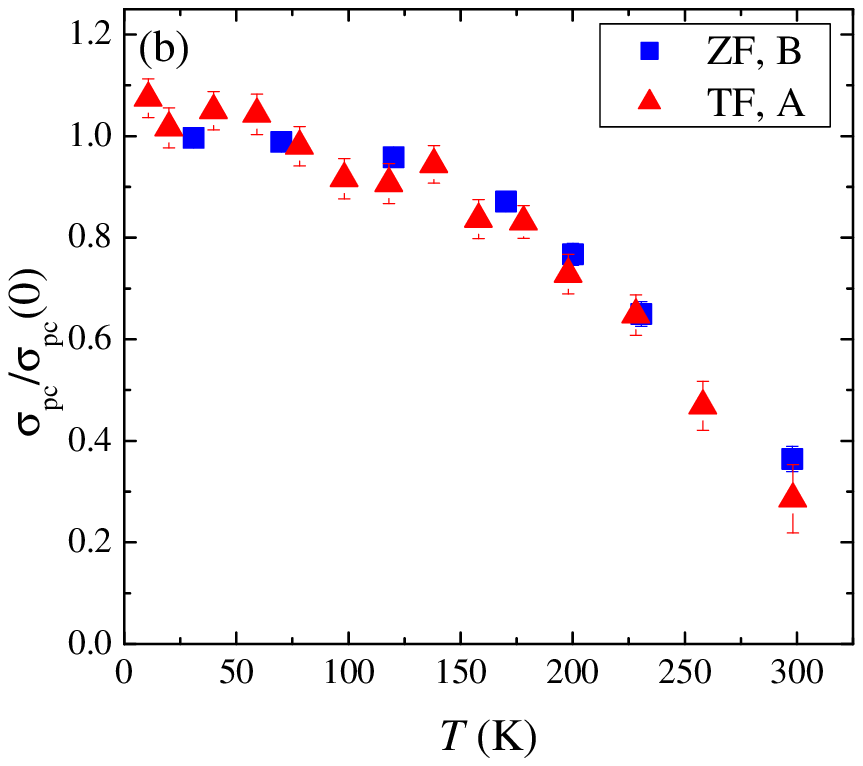}
\caption{(Color online) (a)  Gaussian ($\sigma_{\rm pc}$, full symbols) and exponential
($\lambda_{\rm pc}$, empty symbols)
relaxation rates of MP35N for sample A (circles, triangles) and B (squares) measured in
transverse- (triangles) and zero-field (circles, squares) modes. The solid
horizontal line represents the mean value of $\lambda_{\rm pc}=0.04$ $\mu$s$^{-1}$.
(b) Normalized Gaussian muon relaxation rate $\sigma_{\rm pc}(T)/\sigma_{\rm pc}(0)$
for sample A in the transverse field mode and for
sample B in the zero-field mode.
\label{fig:MP35a}}
\end{figure*}

\begin{figure}[!htb]
\includegraphics[width=0.9\linewidth,trim=0.4cm 0.4cm 0.4cm 0.4cm, clip=true]{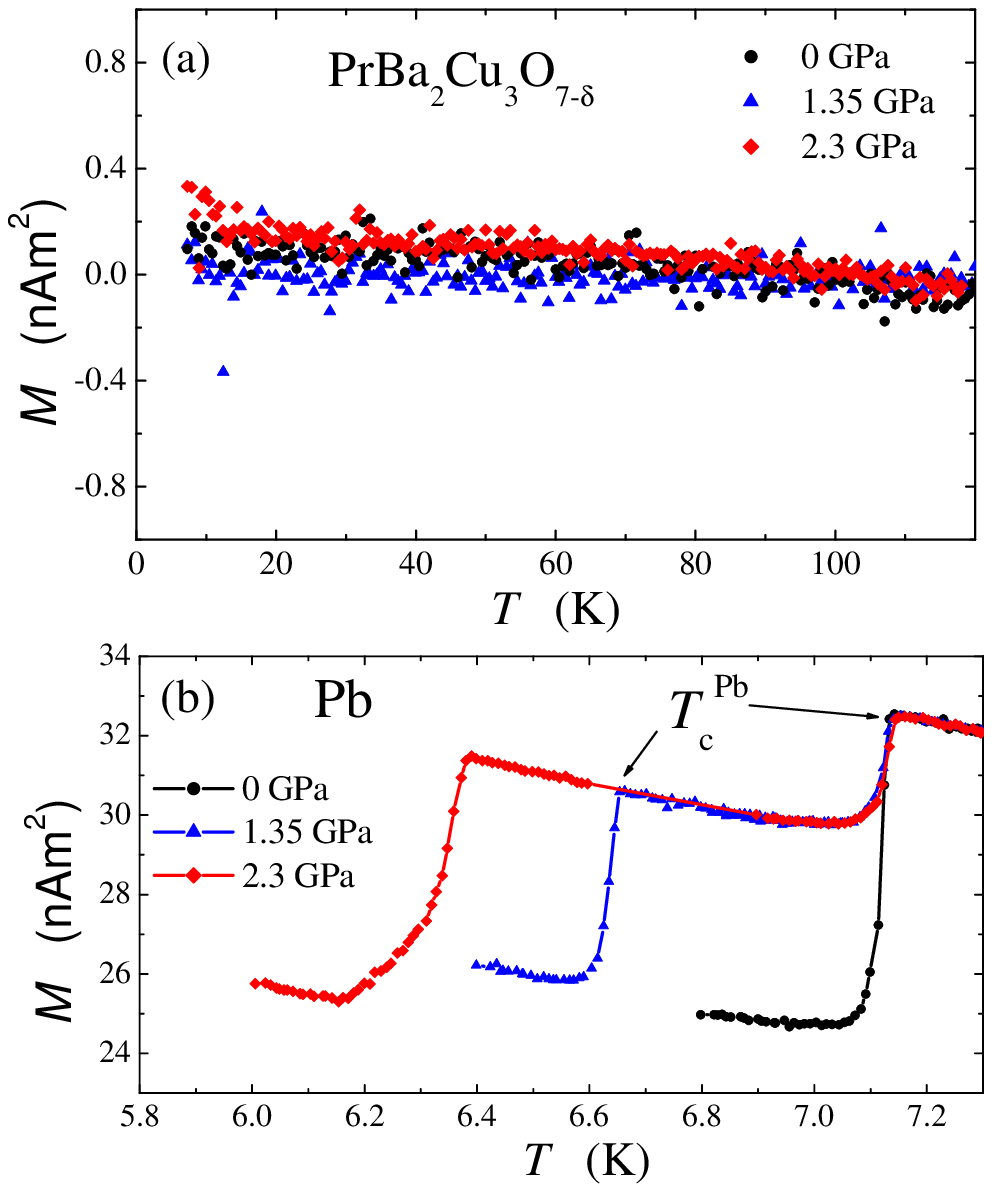}
\caption{(Color online) (a) Field cooled magnetization measurement of 0.5(2) mg of PrBa$_2$Cu$_3$O$_{7-\delta}$
in an applied field of $\mu_0 H=1$ mT. The background magnetization of the pressure cell is subtracted.
(b) Magnetic moment of the pressure cell and the sample in a temperature range below 7.3 K
showing the superconducting transition
temperature of lead pieces inside and outside of the pressure cell (indicated by arrows for $P=1.35$ GPa).
The pressure cell background signal is not subtracted.
\label{fig:MagnzP}}\end{figure}

\section{Conclusions}\label{sec:conclusion}

We investigated the  effect of pressure on the Pr and Cu  N\'eel magnetic
order in Nd$_{1-x}$Pr$_{x}$Ba$_2$Cu$_3$O$_{7-\delta}$ for $x=0.3$, 0.5, 0.7, and 1.
In several aspects the present results differ drastically from those reported previously.\cite{Jee88,Weng99}
The effect of pressure on the Pr  N\'eel ordering temperature $T_{\rm N}^{\rm Pr}$ was found to be
positive for all the investigated Pr concentrations $x$, amounting to $\simeq +0.7$ K/GPa.
This finding is in agreement with the negative effect of pressure on the superconducting transition
temperature $T_{\rm c}$ observed  for $0.1<x<0.3$  and coincides with the result of inelastic
neutron scattering experiments.\cite{Lister99}
This implies a reduction of the effective charge concentration with pressure, resulting in a
suppression of $T_{\rm c}$ and an increase of $T_{\rm N}^{\rm Pr}$ (for $x>0.3$).\cite{Lin00}
In PrBa$_2$Cu$_3$O$_{7-\delta}$
the effect of pressure on the  N\'eel ordering temperature $T_{\rm N}^{\rm Cu}$ for the Cu sublattice  is
weak and nonlinear, suggesting a small pressure dependence of the in- and out-of-plane
exchange integrals for the Cu spins. Thus, the strong effect of
pressure on $T_{\rm N}^{\rm Cu}$ for $0.3<x<0.5$ is dominated
by a reduction of the effective charge concentration in the CuO$_2$ planes with pressure.

\section*{Acknowledgements}
This work was performed at the Swiss Muon Source (S$\mu$S), Paul Scherrer Institut (PSI,
Switzerland). We acknowledge the support by the Swiss National
Science Foundation, the NCCR Materials with Novel
Electronic Properties (MaNEP), the SCOPES grant No.
IZ73Z0-128242, and the Georgian National Science Foundation.

\appendix
\section{ $\mu$SR spectra of the ${\bf \rm Co-Ni}$ alloy (MP35N)}

The nonmagnetic Co-Ni alloy (MP35N)  used for high-pressure $\mu$SR
studies is composed of  35\% Co, 35\% Ni, 20\% Cr, and 10\% Mo (by weight).\cite{AndreicaDiss}
For temperatures above 2 K the parameters of the $\mu$SR spectra may vary
within a range of $\sim$10\%, depending on preparation and aging procedures.
Below 2 K this variation can be more pronounced.
In the temperature range of 2 to 100\,K the $\mu$SR spectra are practically temperature
independent (for details see Ref. \onlinecite{AndreicaDiss}).
Here we present the results for zero and weak transverse field $\mu$SR
of MP35N  in the broad temperature range of 5 to 300 K.
The zero-field $\mu$SR spectra are well described
by Eq. (\ref{eq:PcellZF}).\cite{AndreicaDiss}
For a transverse field $B$, the
muon depolarization may be analyzed with the function:
\begin{equation}\label{eq:PcellTF}
P_{\rm pc, TF}(t) = \cos(\gamma_{\mu}Bt+\phi)\exp
\left(-\frac{1}{2}\sigma_{\rm pc}^2t^2\right)\exp(-\lambda_{\rm pc} t).
\end{equation}
Here, $\sigma_{\rm pc}$ and $\lambda_{\rm pc}$ are the Gaussian and exponential muon
relaxation rates, while $\phi$ is the initial phase of the muon polarization.
A weak transverse field of $B=30$ mT was used for these tests.

Two different samples of MP35N were tested, here referred to as A and B. Both
samples were taken from the same batch, but were subjected to different thermal aging processes.
The parameters $\sigma_{\rm pc}$ and $\lambda_{\rm pc}$ for both samples as
obtained with Eqs. (\ref{eq:PcellZF}) and (\ref{eq:PcellTF}) for
zero and transverse fields in the temperature range of 5 to 300 K are presented in Fig. \ref{fig:MP35a}(a).
The exponential shape of the relaxation rate dependence $\lambda_{\rm pc}$ are identical for both
samples in zero and transverse field
with the temperature independent value $\lambda_{\rm pc}=0.04(1)$ $\mu$s$^{-1}$.
The Gaussian component of the relaxation is nearly temperature
independent below 100 K, in agreement with Ref. \onlinecite{AndreicaDiss}, and decreases
gradually above 100 K. Figure \ref{fig:MP35a}(b) shows the normalized
temperature dependence of $\sigma_{\rm pc}$ for sample A in the transverse-field
mode and for sample B in the zero-field mode.
Although the starting value of $\sigma_{\rm pc}(0)$ is different for both samples,
their normalized temperature dependence is almost identical and not depending on
the mode of measurement (transverse-field or zero-field).

An analysis of the $\mu$SR time spectra of Nd$_{1-x}$Pr$_x$Ba$_2$Cu$_3$O$_{7-\delta}$
in the MP35N pressure cell used in the present study with
Eq. (\ref{eq:AsyTot}) yields  at temperatures below 100 K $\sigma_{\rm pc}(0)=0.31(1)$ $\mu$s$^{-1}$.
The temperature dependence of $\sigma_{\rm pc}$ was obtained from the corresponding
normalized data shown in Fig. \ref{fig:MP35a}(b).
The temperature independent exponential component of the muon relaxation
rate of the pressure cell with $\lambda_{\rm pc}=0.04$ $\mu$s$^{-1}$ was used.

\section{ Magnetization measurements of ${\rm PrBa}_2{\rm Cu}_3{\rm O}_{7-\delta}$}

In order to detect the presence or absence of a superconducting phase with increasing hydrostatic pressure,
magnetization measurements were performed for PrBa$_2$Cu$_3$O$_{7-\delta}$ in field-cooled condition at
an applied field of $\mu_0 H=1$ mT.
The diamond anvil type of pressure cell with a volume under pressure $V_{\rm P}\simeq 0.1$ mm$^3$ was
loaded with PrBa$_2$Cu$_3$O$_{7-\delta}$ sample and a small piece of Pb.
The volume of the sample is about half of $V_{\rm P}$ while the volume of Pb is at least a factor of
$f=5$ smaller than that of sample.
The small piece of Pb inside of $V_{\rm P}$ serves as a pressure manometer
while another piece of Pb outside of $V_{\rm P}$ serves as a reference of the Pb superconducting
transition temperature $T_{\rm c}^{\rm Pb}$ at zero pressure.
Daphne oil was used as a pressure transmission medium.
The results of these measurements for $P=0$, 1.35, and 2.3 GPa are shown in
Fig. \ref{fig:MagnzP}(a). The magnetization of the
pressure cell was measured in a separate experiment and
subtracted. No diamagnetism can be detected within precision of the experiment.
Systematic and statistical errors of the measured
magnetic moments are of the order of 0.2 nAm$^2$.
Figure \ref{fig:MagnzP}(b) shows the magnetization measurements of the superconducting transitions of
lead $T_{\rm c}^{\rm Pb}$s inside and outside of $V_{\rm P}$ at $P=0$, 1.35, and 2.3 GPa.
At $P=2.3$ GPa the transition of Pb is broadened due to a pressure inhomogeneity.
The magnetic moment is also slightly increased, since lead flows around the
nonsuperconducting grains of the sample and increases its effective volume.
The difference of these transition temperatures allows us to determine the applied pressure, while
the drop of the magnetic moment $\delta M^{\rm Pb}\simeq -5$ nAm$^2$
 below $T_{\rm c}^{\rm Pb}$
shows the sensitivity of the measurement.
Namely, for a 100\% superconducting  PrBa$_2$Cu$_3$O$_{7-\delta}$ sample one
expects a drop of the magnetic moment $\Delta M^{\rm Pr}=f\times \delta M^{\rm Pb} = -25$ nAm$^2$
below the respective $T_{\rm c}^{\rm Pr}$. Since we can clearly exclude the presence of a
diamagnetic moment grater than $\delta m=-0.2$ nAm$^2$ a possible superconducting volume fraction is
smaller than $\delta m/\Delta M^{\rm Pr} = 0.2/25=0.008\simeq 1$\%.
Thus, we do not detect the presence of a superconducting
phase in our PrBa$_2$Cu$_3$O$_{7-\delta}$ samples up to 2.3 GPa.


\begin{thebibliography}{99}
\bibitem{comment1}
There are no reports for Pm$_{1-x}$Pr$_x$Ba$_2$Cu$_3$O$_7$
since it is highly radioactive.
\bibitem{commentPch} For a given $x$ the chemical pressure can be increased by reducing
the ionic size of the isovalent rare earth element
$R^{3+}=$ La$^{3+}$, Nd$^{3+}$, Sm$^{3+}$, ..., Yb$^{3+}$, Lu$^{3+}$.
\bibitem{comment3} For a small concentration $x$ ($0< x\ll x_{\rm cr}$) a positive PE on
$T_{\rm c}$ has been observed that is presumably
caused by a transfer of charge from the chain to the plane sites,
also observed in YBa$_2$Cu$_3$O$_{7-\delta}$.\cite{Almasan92,Gupta95,Chen00}
\end{thebibliography}
\end{document}